\documentclass[aps,prb,twocolumn,showpacs,groupedaddress,10pt,a4paper]{revtex4-1}
\usepackage{graphicx}
\usepackage{amsmath}
\usepackage{amssymb}
\usepackage{verbatim}
\usepackage{color}

\newcommand{\m}[1]{\mathrm{#1}}

\usepackage{amsthm}

 \theoremstyle{definition}

 \theoremstyle{remark}

\begin{document}

\title{Quantum memory coupled to cavity modes}
\author{Fabio L. Pedrocchi}
\author{Stefano Chesi}
\author{Daniel Loss}
\affiliation{Department of Physics, University of Basel, Klingelbergstrasse 82, CH-4056 Basel, Switzerland}

\begin{abstract}
Inspired by spin-electric couplings in molecular magnets, we introduce in the Kitaev honeycomb model a linear modification of the Ising interactions due to the presence of quantized cavity fields. This allows to control the properties of the low-energy toric code Hamiltonian, which can serve as a quantum memory, by tuning the physical parameters of the cavity modes, like frequencies, photon occupations, and coupling strengths. We study the properties of the model perturbatively by making use of the Schrieffer-Wolff transformation and show that, depending on the specific setup, the cavity modes can be useful in several ways. They allow to detect the presence of anyons through frequency shifts and to prolong the lifetime of the memory by enhancing the anyon excitation energy or mediating long-range anyon-anyon interactions with tunable sign. We consider both resonant and largely detuned cavity modes.
\end{abstract}


\maketitle
\section{INTRODUCTION}\label{sec:int}

Realizing qubits with long coherence time is a basic requirement for quantum information processing. \cite{DiVincenzo} A promising strategy to achieve this goal is to make use of quantum error correcting codes, where a single logical qubit is represented by suitable entangled states of a collection of many spins. This allows to detect and correct a finite number of errors acting on the individual physical spins of the code. \cite{NielsenChuang} Furthermore, if the quantum code is the ground space of a gapped Hamiltonian, errors in the memory lead to excitations in the spin system and they can be suppressed at low temperature. In this respect, stabilizer Hamiltonians (given by a sum of mutually commuting Pauli operators) are of special importance \cite{Gottesman, NielsenChuang} and a particularly interesting class are stabilizer Hamiltonians with topological order, where the distance of the code grows as a power of the system size $N$ (see Refs. \onlinecite{Bravyi1, Bravyi2} for a rigorous definition of topological quantum order and its consequences on the stability of the ground subspace under local Hamiltonian perturbations). Models with topological order were already known in the context of lattice gauge theory \cite{Wegner, Kogut} and a pioneering example applied to quantum information is the toric code proposed by Kitaev. \cite{KitaevToric}

However, several limitations of this type of model appear when put in contact with an external environment.\cite{Dennis,Castelnovo,Nussinov,Alicki,Pachos,Hamma2009,BravyiNoGo,ColbeckNoGo,ChesiNJP,ChesiPRA,PastawskiPRL,PastawskiLong} Most importantly, it was shown that at any finite temperature the lifetime of the toric code does not grow with system size. \cite{Dennis, Nussinov, Alicki, ChesiPRA} This is a generic feature of stabilizer Hamiltonians with short-range interactions in one and two dimensions since, while the distance of the code increases with $N$, the energy barrier to perform a logical error is bounded by a constant. \cite{BravyiNoGo, ColbeckNoGo} In the case of the toric code, the excitations of the system can be represented as pairs of classical diffusing anyons and the aforementioned energy barrier is simply the energy cost to create a single anyon pair. Therefore, it is of crucial importance to devise new architectures where such gap is enhanced and the anyon population is exponentially suppressed.

It was recently shown that repulsive long-range interactions among the anyons lead to a large self-consistent mean-field gap which grows with system size and the resulting prolongation of the lifetime was studied in detail. \cite{ChesiPRA} The toric code can be realized as a low-energy effective Hamiltonian of the Kitaev honeycomb model, \cite{KitaevHoney, VidalPRB} and long-range interactions between anyons appear in the presence of a non-local coupling with cavity modes extending over the whole memory. \cite{ChesiPRA} In analogy to the spin-electric coupling in molecular magnets, \cite{MirceaPRL, MirceaPRB, CanaliPRB} we consider a modification of the Ising couplings of the type $J_{x,y}\rightarrow J_{x,y}+\delta_{x,y}(a+a^{\dagger})$. In this paper we study in detail the effect of such a coupling in the honeycomb model by making use of perturbation theory and show that cavity modes can be useful in several ways, not only to mediate long-range interactions between anyons.

For example, they allow to realize a basic operation such as the read-out of the error syndrome or, in other words, to detect the presence of anyons across the memory. To achieve this, it is sufficient to couple locally a single cavity mode to the desired read-out site and to detect its frequency shift, similar to a scheme demonstrated for superconducting qubits. \cite{Blais2004,WallraffPRL} As we will see in the next sections, a single cavity mode can also be used to resonantly enhance the anyon gap while, to generate an anyon-anyon interaction of the desired sign, two cavity modes are required. Such a variety of situations corresponds to different specific configurations for the interaction between the cavity modes and the spin model, which we analyze by making use of perturbation theory. Our work is based on the Schrieffer-Wolff transformation, \cite{Phys.Rev.149.491, SWunpublished} which can be easily applied to the Kitaev honeycomb model, even in the presence of cavity modes, to obtain explicit expressions for the parameters of the quantum memory. 

The detailed outline of the paper is as follows. We start with Sec. \ref{sec:model} by introducing the model Hamiltonian and discuss the physical motivation of the coupling. We then describe in Sec. \ref{sec:perturbativeapproach1} the perturbative framework in general terms. In Sec. \ref{sec:readout} we consider different read-out schemes. In Sec. \ref{sec:resonant_enhancement} we study the enhancement of the gap produced by a single resonant cavity mode. In Sec. \ref{sec:long-range} we obtain long-range anyon-anyon interactions from a specific coupling scheme with two resonant cavity modes. We conclude with Sec. \ref{sec_not_resonant} by studying the effect of off-resonant cavity modes and present in Appendices \ref{appendix_SW}-\ref{app:offresonant} an extended discussion of several technical aspects and derivations.

\section{MODEL}\label{sec:model}


\begin{figure}
	\centering
		\includegraphics[width=0.48\textwidth]{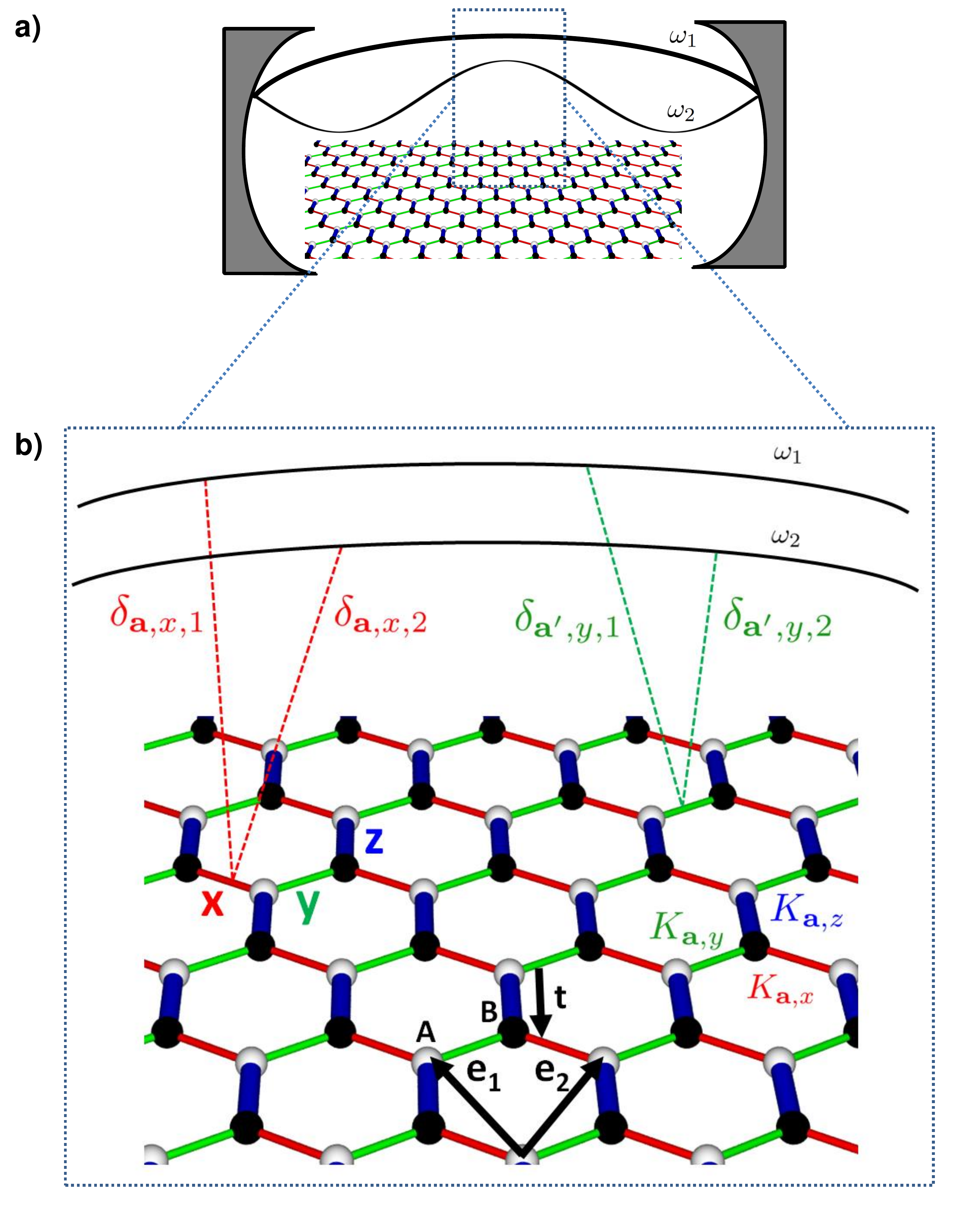}
	\caption{Illustration of the Kitaev honeycomb model with cavity modes. $a)$ Schematic view of the honeycomb model at the center of an optical or microwave cavity. The two shaded areas represent mirrors and the curves in the middle two quantized cavity modes with frequencies $\omega_{1,2}$ to which the spin model couples according to Eq.~(\ref{eq:couplings}). $b)$ Kitaev honeycomb lattice. The spins are represented by white or black dots respectively belonging to the $A$ or $B$ sublattice. Ising interactions of type $\sigma^{x}\sigma^{x}$ along $x$ links (red), $\sigma^{y}\sigma^{y}$ along $y$ links (green), or $\sigma^{z}\sigma^{z}$ along $z$ links (blue) are represented by colored links with different directions. Strongly interacting $z$ dimers are depicted by thick blue links and form a lattice with basis vectors ${\bf e}_{1,2}$ while ${\bf t}$ is a vector which connects $A$- and $B$-sublattice sites. Cavity modes (here, two of them, with frequencies $\omega_{1,2}$) modify the Ising interactions of certain links (as indicated by dashed lines), according to Eq.~(\ref{eq:couplings}).}
	\label{fig:Honeycomb}
\end{figure}

The Kitaev honeycomb model \cite{KitaevHoney} is a two-dimensional version of the quantum compass model defined on a honeycomb lattice. \cite{Kugel, KitaevMott} It consists of $2N$ spin-$1/2$ particles located at the sites of a honeycomb lattice with nearest-neighbor Ising interactions and is illustrated in Fig.~\ref{fig:Honeycomb}b). The interaction between two neighboring spins is either $\sigma^{x}\sigma^{x}$, $\sigma^{y}\sigma^{y}$, or $\sigma^{z}\sigma^{z}$, depending on the direction of the link connecting the two spins. In this work we are interested in the limit when the $\sigma^{z}\sigma^{z}$ couplings are much stronger than the other ones. Therefore, the model can be seen as consisting of $N$ weakly coupled dimers, forming a triangular lattice with unit vectors ${\bf e}_{1,2}$, as shown in Fig.~\ref{fig:Honeycomb}b). 

In this work, we consider the following generalization of the Kitaev honeycomb Hamiltonian
\begin{eqnarray}\label{eq:kitaevhoneycomb}
H &=&\sum_{\alpha}\omega_{\alpha}a_{\alpha}^{\dagger}a_{\alpha} -\sum_{{\bf a}}K_{{\bf a},z}\sigma_{{\bf a}}^{z}\sigma_{{\bf a}+{\bf t}}^{z}\nonumber\\
&&-\sum_{{\bf a}}K_{{\bf a},x}\sigma_{{\bf a}}^{x}\sigma_{{\bf a}+{\bf e}_{1}+{\bf t}}^{x}-\sum_{{\bf a}}K_{{\bf a},y}\sigma_{{\bf a}}^{y}\sigma_{{\bf a}+{\bf e}_{2}+{\bf t}}^{y},
\end{eqnarray}
where the three last sums run over all vectors ${\bf a}=n_{1}{\bf e}_1+n_{2}{\bf e}_{2}$ ($n_{1,2}\in\mathbb{Z}$) pointing to a $A$-sublattice site (white dots), $\bf t$ is a vector which connects $A$- (white dots) and $B$- (black dots) sublattice sites and $\sigma^{x,\,y,\,z}$ are the usual spin Pauli operators. In Eq.~(\ref{eq:kitaevhoneycomb}), we have assumed that the Ising couplings $K_{{\bf a},k}$ (with $k=x,y,z$) are site dependent. Furthermore, we introduced a collection of cavity modes labeled by $\alpha$, with frequency $\omega_{\alpha}$, and annihilation operators  $a_{\alpha}$. We propose that they are coupled to the spin model as follows:
\begin{eqnarray}\label{eq:couplings}
K_{{\bf a},k}= J_{{\bf a},k}+\sum_{\alpha}\delta_{{\bf a},k,\alpha}(a_{\alpha}+a_{\alpha}^{\dagger}),
\end{eqnarray}
where $J_{{\bf a},k}$ are the unperturbed Ising couplings, and $\delta_{{\bf a},k,\alpha}$ describe the linear change due to the cavity modes (see Fig.~\ref{fig:Honeycomb}). The original Kitaev model is recovered for $\delta_{{\bf a},k,\alpha}=0$ and $J_{{\bf a},k}=J_k$, independent of the lattice site. For simplicity, we also assume in the rest of this work $K_z= J_z$, but we keep the more general form of coupling [Eq.~(\ref{eq:couplings})] for $K_{{\bf a},x}$ and $K_{{\bf a},y}$. 

To physically justify Eq.~(\ref{eq:couplings}), we note that Ising couplings can be engineered in a variety of systems, for example between  quantum dots, \cite{MirceaCoulomb} Josephson Junction qubits, \cite{PashkinNature, AverinBruder, Nori}, atoms in optical lattices, \cite{MuellerNaturePhysics, LukinPRL} and doped coupled cavities.\cite{Angelakis} Anisotropic spin interactions are also present in molecular magnets \cite{MirceaPRL,MirceaPRB, CanaliPRB} and between pseudospin orbital states. \cite{Kugel} Recently, theoretical proposals to engineer the honeycomb model in Mott insulators with strong spin-orbit coupling were discussed in Refs. \onlinecite{KitaevMott, PRLChaloupka}. Metal-oxide compounds (layered iridates of type $A_{2}\m{Ir} \m{O}_{3}$, with $A=\m{Li}, \m{Na}$) were considered there as promising materials. In these systems, the strength of the magnetic interaction can be modified by external perturbations, in particular electric and magnetic fields. \cite{AverinBruder,MirceaCoulomb,MirceaPRL, MirceaPRB, CanaliPRB} Such perturbations can be generated by charged nanomechanical systems (e.g., cantilevers) or quantized electromagnetic fields (e.g., in cavities and transmission line resonators \cite{Blais2004, WallraffPRL, Burkard}), thus realizing a coupling of the type of Eq.~(\ref{eq:couplings}).
\begin{figure}
	\centering
		\includegraphics[width=0.45\textwidth]{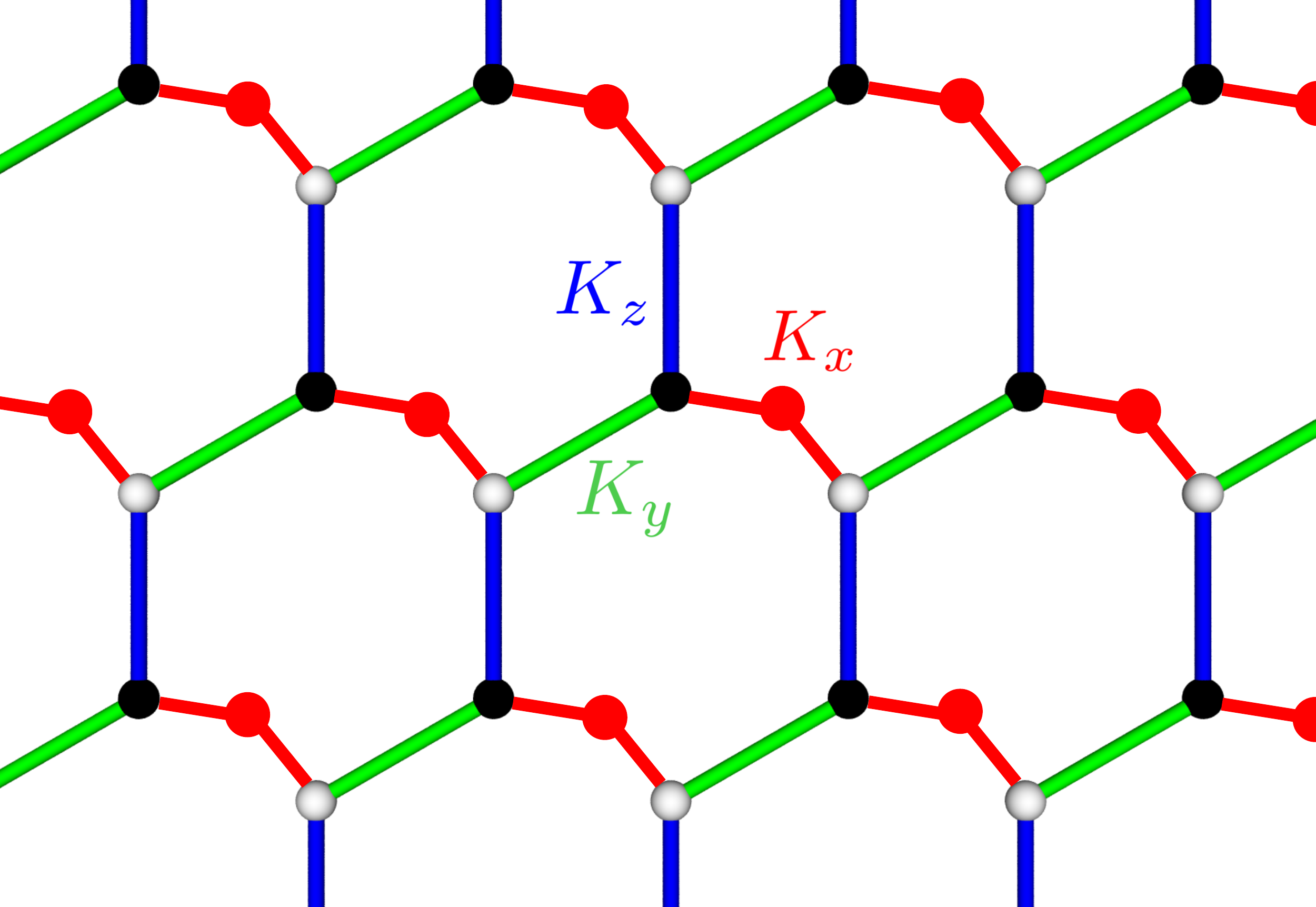}
	\caption{Pictorial representation of superexchange interactions. The bridge sites (dots in the middle of each $x$-link) mediate the magnetic $\sigma^{x}\sigma^{x}$ interactions between the spins (dots at each site of the honeycomb lattice). Here, the $x$ links have a nonvanishing dipole moment and can thus be linearly coupled to external electric fields. By controlling the equilibrium position of the bridge sites, the coupling to the cavity modes can in principle be tuned locally.}
	\label{fig:SuperExchange}
\end{figure}
As a more specific example, we would like to mention molecular magnets. In this case, electric fields can be used to modify the overlap of the orbital wave functions and thus the resulting magnetic interactions, which are determined by exchange or superexchange mechanisms. Strongly anisotropic (Ising-like) magnetic interactions can arise here in the presence of spin-orbit coupling. \cite{MirceaPRL,MirceaPRB, CanaliPRB}

A linear effect in the electric field, as in Eq.~(\ref{eq:couplings}), can only exist if the electric-dipole of the underlying bond is not zero or, in other terms, if the inversion symmetry of that bond is broken. This is illustrated in Fig.~\ref{fig:SuperExchange} for the $x$ links, assuming the presence of superexchange interactions mediated via an auxiliary site (the so-called bridge site). Figure~\ref{fig:SuperExchange} also illustrates that only selected links can be coupled to the cavity modes since $y$- and $z$-bonds have no dipole moment and thus $\delta_{{\bf a},k,\alpha}=0$. Finally, Fig.~\ref{fig:SuperExchange} suggests that $\delta_{{\bf a},k,\alpha}$ can be controlled locally. By realizing Fig.~\ref{fig:SuperExchange} in a quantum dot setup, $\delta_{{\bf a},k,\alpha}$ could be made zero or not by changing the equilibrium position of the bridge sites with static electric gates.

\section{PERTURBATIVE APPROACH}\label{sec:perturbativeapproach1}

\begin{figure}
	\centering
		\includegraphics[width=0.45\textwidth]{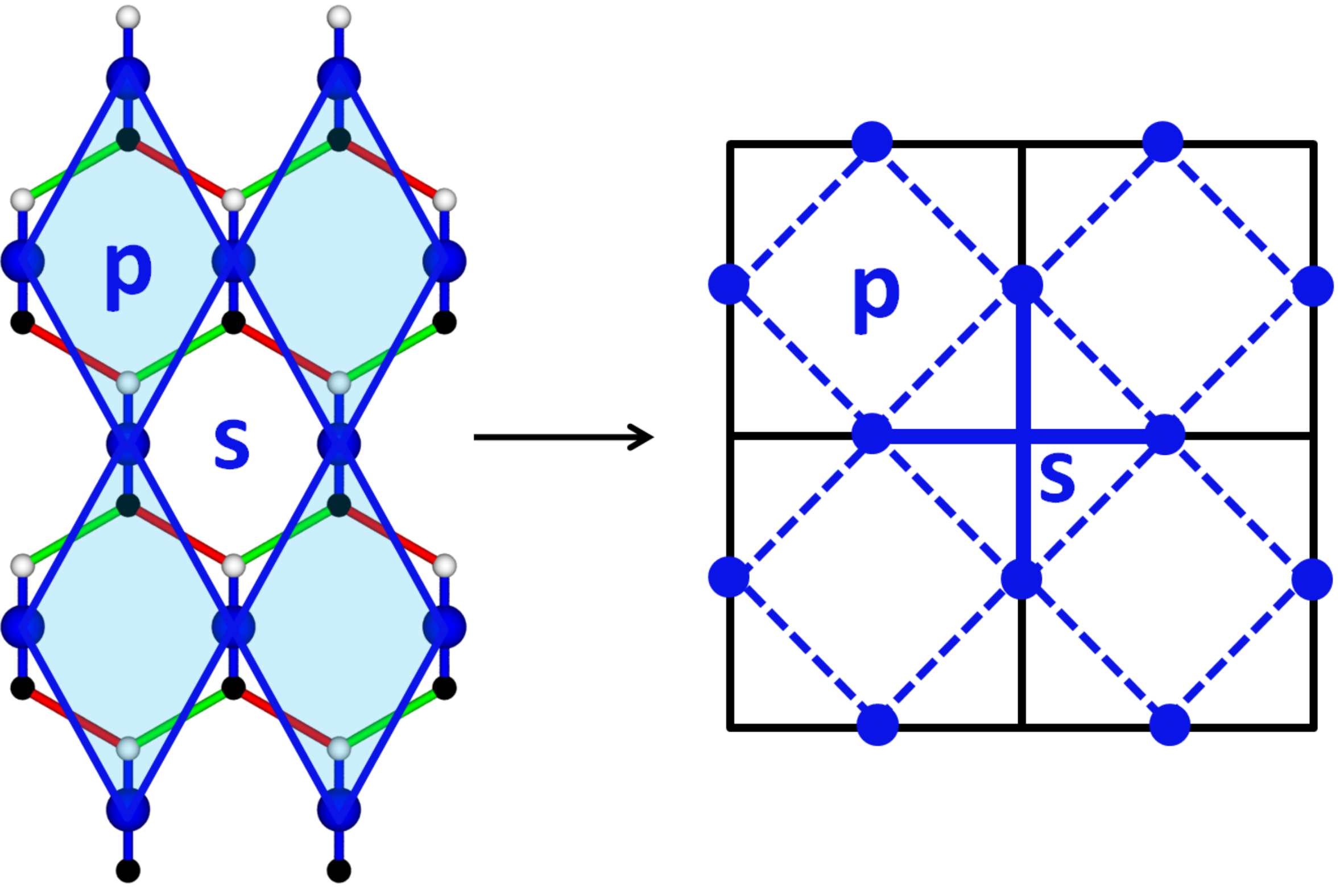}
	\caption{Mapping of the Kitaev honeycomb model to the toric code. Left-hand side: each $z$ dimer, formed by a black and a white spin [see also Fig.~\ref{fig:Honeycomb}b)], corresponds to a new lattice site (large blue dots). To these new sites, hard-core bosons and effective spins are associated. A low-energy (zero hard-core bosons) toric code Hamiltonian for the effective spins is obtained by perturbation theory. The resulting toric code model is schematically illustrated on the right-hand side. The dark and white unit cells on the left-hand-side correspond now to plaquette (p) and star (s) operators, respectively.}
	\label{fig:HardCoreBoson}
\end{figure}

Exact results for the Kitaev honeycomb model exist for arbitrary values of $J_{x,y,z} $ \cite{KitaevHoney} but they do not include couplings to the cavity modes. Therefore, we study Eq.~(\ref{eq:kitaevhoneycomb}) with perturbation theory. This is indeed appropriate if the model is intended to be a quantum memory, since for $J_{x,y} \ll J_z$ the honeycomb model can be mapped to the toric code. \cite{KitaevHoney,VidalPRB} Without cavity modes, perturbative treatments of the honeycomb model can be found in Ref. \onlinecite{KitaevHoney} where the author makes use of Green's functions, and more recently in Ref. \onlinecite{VidalPRB} with the perturbative continuous unitary transformation method. \cite{WegnerPCUT,UhrigPCUT} This work is based on a different approach, namely on the Schrieffer-Wolff (SW) transformation, \cite{Phys.Rev.149.491,SWunpublished} which can be applied in a straightforward way to Eq.~(\ref{eq:kitaevhoneycomb}), as we will see in the next sections.

\subsection{Hard-core boson transformation}

The first step of our analysis consists in the hard-core boson transformation presented in Ref. \onlinecite{VidalPRB}, which maps each $z$ dimer to a particle and an effective spin, see Fig.~\ref{fig:HardCoreBoson}. The two spins of each $z$-dimer can be either parallel or antiparallel and, if we assume that $J_{z}>0$, the parallel configuration has lower energy than the antiparallel one. The excited dimers can thus be in	terpreted as particles (hard-core bosons) with energy $2J_{z}$, and the remaining degree of freedom as an effective spin-$1/2$. More explicitly,
\begin{eqnarray}\label{hcb_states}
&\vert\uparrow\rangle_{\bf a}\otimes\vert\uparrow\rangle_{{\bf a}+{\bf t}},=\vert\Uparrow0\rangle_{\bf a},\qquad
&\vert\downarrow\rangle_{\bf a}\otimes\vert\downarrow\rangle_{{\bf a}+{\bf t}}=\vert\Downarrow0\rangle_{\bf a}, \nonumber \\
&\vert\downarrow\rangle_{\bf a}\otimes\vert\uparrow\rangle_{{\bf a}+{\bf t}}=\vert\Uparrow1\rangle_{\bf a},\qquad
&\vert\uparrow\rangle_{\bf a}\otimes\vert\downarrow\rangle_{{\bf a}+{\bf t}},=\vert\Downarrow1\rangle_{\bf a}, \nonumber
\end{eqnarray}
where the four possible spin configurations of a given dimer (left-hand sides) are mapped to states with $0,1$  hard-core boson and $\Downarrow,\Uparrow$ effective spin. If we rewrite the above transformation in operator language we obtain
\begin{equation}\label{eq:hard-core}
\begin{array}{ll}
\sigma_{{\bf a}+{\bf t}}^{x} =\tau_{{\bf a}}^{x}(b_{{\bf a}}^{\dagger}+b_{{\bf a}}),  \quad &\sigma_{\bf a}^{x}=b_{\bf a}^{\dagger}+b_{\bf a},\\
\sigma_{{\bf a}+{\bf t}}^{y} =\tau_{{\bf a}}^{y}(b_{{\bf a}}^{\dagger}+b_{{\bf a}}),  &\sigma_{\bf a}^{y}=i\tau_{\bf a}^{z}(b_{\bf a}^{\dagger}-b_{\bf a}),\\
\sigma_{{\bf a}+{\bf t}}^{z} =\tau_{{\bf a}}^{z},   &\sigma_{\bf a}^{z}=\tau_{\bf a}^{z}(1-2b_{\bf a}^{\dagger}b_{\bf a}),
\end{array}                      
\end{equation}
where $b_{\bf a}^{\dagger}\,(b_{\bf a})$ creates (destroys) a hard-core boson and $\tau_{\bf a}^{x,y,z}$ are the Pauli operators of the effective spin-$1/2$ at site $\bf a$. At distinct sites $\bf a$ and $\bf a'$, bosonic commutation relations are satisfied:
\begin{equation}
\qquad [b_{\bf a},b_{\bf a'}^{\dagger}]=0, \qquad {\rm if}~\bf a\neq \bf a'.
\end{equation}
Furthermore, $\{b_{\bf a},b_{\bf a}^{\dagger }\}=1$.

By making use of these bosonic operators, we can rewrite the first line of Eq.~(\ref{eq:kitaevhoneycomb}) simply as
\begin{equation}\label{H0}
H_{0}=\sum_{\alpha}\omega_{\alpha}a_{\alpha}^{\dagger}a_{\alpha}+2 J_z \sum_{\bf a} b^\dag_{\bf a} b_{\bf a},
\end{equation}
which represents the unperturbed Hamiltonian. Here and in the rest of the paper we drop an inessential constant $-J_z N$.

\subsection{Perturbation operators}

We now consider the second line of Eq.~(\ref{eq:kitaevhoneycomb}), which constitutes our perturbation. By rewriting the spin operators by making use of Eq.~(\ref{eq:hard-core}), we obtain that Eq.~(\ref{eq:kitaevhoneycomb}) takes the following form 
\begin{eqnarray}\label{eq:SWTH}
H&=&H_{0}+T_{0}+T_{-2}+T_{+2},
\end{eqnarray}
where
\begin{eqnarray}
\label{eq:operatorsnophoton1}
&& T_{0}=-\sum_{{\bf a},k}(K_{{\bf a},k} t_{{\bf a},k}+\mathrm{H.c.}):= f_0(K_{{\bf a},k}),\\
\label{eq:operatorsnophoton2}
&& T_{+2}=-\sum_{{\bf a},k} K_{{\bf a},k} v_{{\bf a},k}:= f_{+2}(K_{{\bf a},k}),\\
&& T_{-2}=\left(T_{+2}\right)^{\dagger}:= f_{-2}(K_{{\bf a},k})\label{eq:operatorsnophoton3},
\end{eqnarray}
with $k=x,y$. As in Ref. \onlinecite{VidalPRB}, we defined the hopping operators
\begin{eqnarray}
t_{{\bf a},x}&=&b_{{\bf a}+{\bf e}_{1}}^{\dagger}b_{{\bf a}}\tau_{{\bf a}+{\bf e}_{1}}^{x},\\ 
t_{{\bf a},y}&=&-ib^\dag_{{\bf a}+{\bf e}_{2}}b_{{\bf a}}\tau_{{\bf a}+{\bf e}_{2}}^{y}\tau_{{\bf a}}^{z},
\end{eqnarray}
and the hard-core boson creation operators
\begin{eqnarray}
v_{{\bf a},x}&=&b_{{\bf a}+{\bf e}_{1}}^{\dagger}b_{{\bf a}}^{\dagger}\tau_{{\bf a}+{\bf e}_{1}}^{x},\\ 
v_{{\bf a},y}&=&ib_{{\bf a}+{\bf e}_{2}}^{\dagger}b_{{\bf a}}^{\dagger}\tau_{{\bf a}+{\bf e}_{2}}^{y}\tau_{{\bf a}}^{z}.
\end{eqnarray}
In other words, $T_{+2(-2)}$ creates (destroys) two nearest-neighbor hard-core bosons and $T_{0}$ makes a hard-core boson hop to a nearest-neighbor unoccupied dimer.

It is also useful to keep track of the change in the number of photons: whenever a hard-core boson is created, destroyed, or hops from one site to another, a photon might be simultaneously created or destroyed. Therefore, we introduce $T^{m\alpha}_n$ operators, where the lower label ($n=0,\pm2$, as before) refers to the change in the number of hard-core bosons, while the upper label ($m=0,\pm1$) indicates that the number of photons in mode $\alpha$ changes by $m$. Such operators are simply defined by substituting in Eqs.~(\ref{eq:operatorsnophoton1}-\ref{eq:operatorsnophoton3}) the full couplings $K_{{\bf a},k}$ by the appropriate quantity, $J_{{\bf a},k}$, $\delta_{{\bf a},k,\alpha}a_\alpha$, or $\delta_{{\bf a},k,\alpha}a_\alpha^{\dag}$. By using the notation introduced in Eqs.~(\ref{eq:operatorsnophoton1}-\ref{eq:operatorsnophoton3}), which define the $f_n$ functions, we can write
\begin{eqnarray}\label{eq:operators}
&& T^0_{n}= f_n(J_{{\bf a},k}), \\
&& T_{n}^{+\alpha}= f_n(\delta_{{\bf a},k,\alpha}) a_\alpha^\dag ,\\ 
&& T_{n}^{-\alpha}= f_n(\delta_{{\bf a},k,\alpha})  a_\alpha. 
\end{eqnarray}
Clearly, the $T_n$ is also given by a sum of the $T^{m\alpha}_n$:
\begin{equation}\label{Tn_sum}
T_n = T_n^0+\sum_{\alpha,\pm} T_{n}^{\pm\alpha}.
\end{equation}
Furthermore, the energy change corresponding to $T_n^{m\alpha}$ is immediately obtained from the energies $2J_z$ and $\omega_\alpha$ of the hard-core bosons and the photons respectively, and can be expressed through the following commutation relations
\begin{equation}\label{commutations2}
\left[H_{0},T_{n}^{m\alpha}\right]= (2 n J_z + m \omega_\alpha) T_{n}^{m\alpha}.
\end{equation}
Note that we use here the convention $0\alpha \equiv 0$ such that $T^0_0$ and $T^0_{\pm 2}$, which do not change the state of the cavity, are defined independently of the value of $\alpha$. 

\subsection{SW transformation and correspondence to the toric code}\label{sec:toric_correspondence}

In the perturbative limit we are interested in, the SW transformation allows to obtain an effective Hamiltonian in a desired subspace up to an arbitrary order in perturbation theory. For the convenience of the reader, we summarize the general procedure in Appendix~\ref{appendix_SW}, and provide there the general formulas appropriate for the type of Hamiltonian of interest in this work, up to the fourth perturbative order. Since the quantum information is encoded in the spin degrees of freedom, we always consider the low-energy subspace where no hard-core boson is present. On the other hand, we will generally allow the modes of the cavity to be excited, to study how the presence of a finite photon population affects the properties of the memory. While the treatment of Appendix \ref{appendix_SW} is completely general, the resulting fourth-order expressions involve too many terms to be presented here. It is instead interesting to consider specific coupling schemes and physically relevant regimes, in which only a few dominant contributions are important. Several examples will be examined in detail in the following sections. 

In the remainder of this section we restrict ourselves to the case without cavity modes, and derive the toric code from the formulas of Appendix~\ref{appendix_SW}. By setting all the $\delta_{{\bf a},k,\alpha}$ to zero, the $T_n$ operators coincide with $T_n^0$ and all the summations on the photon indexes $i,j,k,r$ in Eqs.~(\ref{SW_2}-\ref{SW_4}) can be dropped. By applying such formulas in the subspace with zero hard-core bosons we obtain 
\begin{eqnarray}\label{eq:SWT_no_cavity}
H_{\m{eff}}=&& -\frac{1}{4J_{z}}T_{-2}T_{+2}+\frac{1}{16J_{z}^2}T_{-2}T_{0}T_{+2}\nonumber\\
&&-\frac{1}{128J_{z}^3}T_{-2}T_{-2}T_{+2}T_{+2}-\frac{1}{64J_{z}^3}T_{-2}T_{0}T_{0}T_{+2}\nonumber\\
&&+\frac{1}{64J_{z}^3}T_{-2}T_{+2}T_{-2}T_{+2},
\end{eqnarray}
where we wrote explicitly the third-order term in the second line, even if it gives no contribution: a hard-core-boson pair created by $T_{+2}$ from the vacuum, followed by a $T_0$ hopping process, cannot be annihilated by the $T_{-2}$ operator. In general, only even orders appear in the perturbation theory and Eq.~(\ref{SW_3}) always evaluates to zero. It is also worth pointing out that Eq.~(\ref{eq:SWT_no_cavity}) is the same as the one derived in Ref. \onlinecite{VidalPRB} with the perturbative continuous unitary transformation approach.

\begin{figure}
	\centering
		\includegraphics[width=0.40\textwidth]{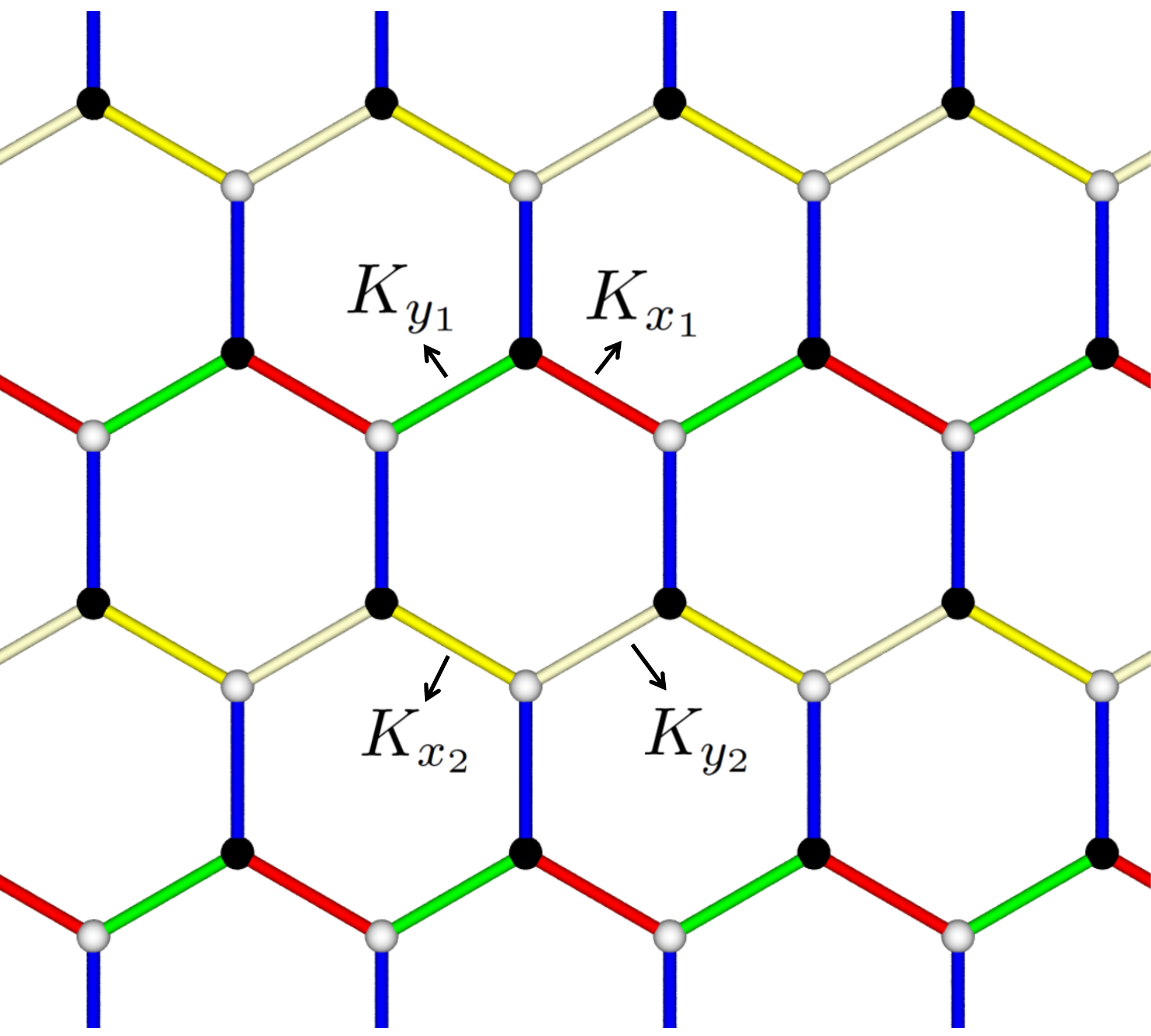}
	\caption{Honeycomb lattice with two distinct $x$-couplings $K_{x_{1,2}}$ and two distinct $y$-couplings $K_{y_{1,2}}$.}
	\label{fig:differentlinks}
\end{figure}

If we now explicitly write Eq.~(\ref{eq:SWT_no_cavity}) in terms of spin operators, by using Eqs.~(\ref{eq:operatorsnophoton1}-\ref{eq:operatorsnophoton3}) with $K_{{\bf a},k}=J_{k}$, the toric code Hamiltonain is obtained. \cite{KitaevToric} As a slight generalization, we consider here the case when four distinct couplings enter, namely $J_{x_{1,2}}$ and $J_{y_{1,2}}$. This specific scheme is illustrated in Fig.~\ref{fig:differentlinks} and leads to
\begin{eqnarray}\label{eq:SWTtwocouplings}
&&H_{\m{eff}}=-\frac{J_{x_{1}}^2+J_{x_{2}}^2}{4J_{z}}\frac{N}{2}-\frac{J_{y_{1}}^2+J_{y_{2}}^2}{4J_{z}}\frac{N}{2}\nonumber\\
&&-\frac{J_{x_{1}}^2J_{y_{1}}^2+J_{x_{2}}^2J_{y_{2}}^2}{16J_{z}^3}\frac{N}{2}+\frac{J_{x_{1}}^2J_{y_{2}}^2+J_{x_{2}}^2J_{y_{1}}^2}{16J_{z}^3}\frac{N}{2}\nonumber\\
&&-\frac{J_{x_{1}}^2J_{x_{2}}^2+J_{y_{1}}^2J_{y_{2}}^2}{16J_{z}^3}\frac{N}{2}+\frac{J_{x_{1}}^4+J_{x_{2}}^4}{64J_{z}^3}\frac{N}{2}+\frac{J_{y_{1}}^4+J_{y_{2}}^4}{64J_{z}^3}\frac{N}{2}\nonumber\\
&&-\frac{J_{x_{1}}J_{y_{1}}J_{x_{2}}J_{y_{2}}}{16J_{z}^3}\sum_{{\bf a}}W_{{\bf a}}.
\end{eqnarray}
This result will be useful later, when in Sec.~\ref{sec_not_resonant} we will discuss the non-resonant coupling of the cavity modes to the spin Hamiltonian. The first three lines are a constant and can thus be dropped for the moment. We have kept them here since they will become important in Sec.~\ref{sec_not_resonant}, when the $J_{x_{1,2}}$, $J_{y_{1,2}}$ will be generalized to full couplings $K_{x_{1,2}}$, $K_{y_{1,2}}$, including the cavity modes.

The last line of Eq.~(\ref{eq:SWTtwocouplings}) is expressed in terms of the  spin operators $W_{\bf a}$, defined as
\begin{equation}\label{eq:Wi}
W_{\bf a} = \tau_{{\bf a}}^{z}\tau_{{\bf a}+{\bf e}_2}^{y}\tau_{{\bf a}+{\bf e}_1}^{y}\tau_{{\bf a}+{\bf e}_1+{\bf e}_2}^{z}.
\end{equation}
Note that the $W_{\bf a}$ operator involves four dimers, connected by two $x$ links and two $y$ links. We will use sometimes the expression that such links ``belong to'' the corresponding operator $W_{\bf a}$. The $W_{\bf a}$ operators are mutually commuting and can be rewritten in terms of stars and plaquettes of the toric code, as introduced in Ref. \onlinecite{KitaevToric}. In order to see this correspondence, we perform a spin rotation. \cite{VidalPRB} If the spin lies at a site which is the bottom corner of a dark unit cell (see the left-hand side of Fig.~\ref{fig:HardCoreBoson})
\begin{equation}
\tau^{x}=-s^{y},~~~  \tau^{y}=s^{x}, ~~~ \tau^{z}=s^{z},
\end{equation}
otherwise
\begin{equation}
\tau^{x}=s^{y}, ~~~ \tau^{y}=s^{z}, ~~~ \tau^{z}=s^{x}.
\end{equation}
We have now that half of the unit cells of the lattice of dimers (the dark unit cells in the left-hand side of Fig.~\ref{fig:HardCoreBoson}) correspond to plaquette operators $B_p$ of the form $s^{z}s^{z}s^{z}s^{z}$. The other half of the cells (the empty ones in the left-hand side of Fig.~\ref{fig:HardCoreBoson}) correspond to star operators $A_s$ of the form $s^{x}s^{x}s^{x}s^{x}$. By setting $J_{x_{1}}=J_{x_{2}}=J_{x}$ and $J_{y_{1}}=J_{y_{2}}=J_{y}$, Eq.~(\ref{eq:SWTtwocouplings}) gives back the result of Refs. \onlinecite{KitaevToric, VidalPRB} 
\begin{eqnarray}\label{eq:toric}
H_{\m{eff}}=&&-\frac{J_{x}^2+J_{y}^2}{4J_{z}}N-\frac{J_{x}^4+J_{y}^4}{64J_{z}^3}N \nonumber\\
&& -\frac{J_{x}^2J_{y}^2}{16J_{z}^3}\left(\sum_{s}A_{s}+\sum_{p}B_{p}\right).
\end{eqnarray}
In the following we will use the simpler notation in terms of the $W_{\bf a}$ operators, by keeping in mind that
\begin{equation}\label{eq:Wi_sum}
\sum_{\bf a} W_{\bf a} =\sum_{s}A_{s}+\sum_{p}B_{p}.
\end{equation}

\section{READ-OUT SCHEMES}\label{sec:readout}

As a first application of the coupling of the honeycomb spin model to cavity modes, we show that it allows the read-out of the eigenvalues of the star and plaquette operators. This is a basic requirement if the toric code model is aimed to be a quantum memory, since the knowledge of such eigenvalues (the error syndrome) is needed to perform the error-correction algorithm and to retrieve the encoded information. 

We start by discussing an example when a single cavity mode, with unperturbed frequency $\omega$ and annihilation operator $a$, allows to read the eigenvalue of a single operator $W_{{\bf a}_r}$, for a given site ${\bf a}_r$. The specific coupling scheme we consider for Eq.~(\ref{eq:couplings}) is as follows, for the $x$ links:
\begin{equation}\label{couplings_single_readout}
K_{{\bf a}_r,x}=K_{{\bf a}_r+{\bf e}_{2},x}= J_x + \delta_x (a+a^\dag), 
\end{equation}
and $K_{{\bf a},x}=J_x$ otherwise. Furthermore, $K_{{\bf a},y}= J_y$ for all the $y$ links. In other terms, only the two $x$ links which belong to the $W_{{\bf a}_r}$ operator are coupled to the cavity mode. This allows to read out the eigenvalue of $W_{{\bf a}_r}$ from an indirect measurement since we obtain the following effective Hamiltonian:
\begin{equation}\label{eq:readout}
H_{\rm eff}=\left[\omega-\frac{2\delta_{x}^2}{4J_{z}-\omega} -\frac{4\delta_{x}^2J_{y}^2}{(4J_{z}-\omega)^3} W_{{\bf a}_{r}}\right] a^{\dagger}a,
\end{equation}
which is derived from Eq.~(\ref{eq:kitaevhoneycomb}) by applying the general formalism of Appendix~\ref{appendix_SW}. In particular, as described in more detail in Appendix~\ref{app:SWONEMODE}, we have evaluated Eqs.~(\ref{SW_2}) and (\ref{SW_4}) for the specific coupling scheme under consideration, keeping only the terms diagonal in the number of photons and assuming the resonant condition
\begin{equation}\label{resonance_cond}
 (4J_{z}-\omega) \ll 4J_{z},\omega.
\end{equation}
This implies that, at each order, only resonant terms contribute to Eq.~(\ref{eq:readout}), i.e., the ones where only powers of $( 4J_{z}-\omega )$ appear in the energy denominators. Here, and in the rest of this work, we also assume $(4J_z -\omega)>0$, such that the state with zero hard-core bosons is the unperturbed ground state of the resonant subspace.

Equation~(\ref{eq:readout}) shows that the frequency of the cavity is now
\begin{equation}
\omega-\frac{2\delta_{x}^2}{4J_{z}-\omega}\mp \frac{4\delta_{x}^2J_{y}^2}{(4J_{z}-\omega)^3}.
\end{equation}
Therefore, it is sufficient to detect the change in the frequency of the cavity mode, which is conditioned by the eigenvalue of $W_{{\bf a}_{r}}$($\pm1$). Let us also recall from Sec.~\ref{sec:toric_correspondence} that the lattice site ${\bf a}_{r}$ determines if the operator $W_{{\bf a}_{r}}$, defined in Eq.~(\ref{eq:Wi}), corresponds to a star (empty unit cells in the left hand side of Fig.~\ref{fig:HardCoreBoson}) or a plaquette (dark unit cells in the left hand side of Fig.~\ref{fig:HardCoreBoson}) of the toric code model. 

It is worth mentioning that a similar read-out scheme was demonstrated in superconducting circuits for a single qubit: in that case the role of $W_{{\bf a}_r}$ is played by $\sigma_z$ of the qubit being read, dispersively coupled to a transmission line resonator. \cite{Blais2004, WallraffPRL} The frequency shift of the cavity mode can be determined in transmission measurements. Indeed, proposals for the realization of the honeycomb spin Hamiltonian with superconducting circuits already exist, \cite{Nori} and it is conceivable that our readout scheme could also be realized in this type of systems. Note that the read-out must take place in a time interval which is much smaller than the typical time for a spin to flip. This condition can be satisfied for a single superconducting qubit, since the relaxation time can reach about $10~\mu$s, while the response time of the measurement is a few hundred nanoseconds. \cite{WallraffPRL} However, the signal to noise ratio is presently too low to allow single-shot read-out.
Besides superconducting systems, also spin qubits in semiconductor nanostructures can be coupled to transmission line resonators via spin-orbit interactions. In particular, InAs nanowire quantum dots were discussed in Ref. \onlinecite{MirceaInAs}.

We now address the problem of measuring all the stars and plaquettes of the memory. As discussed already at the end of Sec.~\ref{sec:model}, we can imagine that all the couplings $\delta_{{\bf a},x,\alpha}$ can be switched on/off. For our read-out scheme, this implies that the site ${\bf a}_r$ can be freely chosen and all plaquettes and stars can be successively measured, one after the other. This method is not efficient in terms of the total read-out time for the whole memory, but has the advantage of using a single cavity mode. Having at disposal many cavity modes (ideally, one for each site of the memory), much more efficient schemes can be realized. 

Consider in particular a set of sites $\Omega = \{ {\bf a}_\alpha \}$, each of them corresponding to an operator $W_{{\bf a}_\alpha}$  which we would like to measure with a cavity mode $\alpha$. For simplicity, we assume that all frequencies and couplings are equal, which can be realized with an array of identical resonators. Each mode $\alpha$ (in particular, the read-out of its frequency) can be addressed individually and is coupled to the two $x$ links of the corresponding $W_{{\bf a}_\alpha}$ operator in the following way
\begin{equation}\label{couplings_multiple_readout}
K_{{\bf a}_\alpha,x}=K_{{\bf a}_\alpha+{\bf e}_2,x}= J_x + \delta_x (a_\alpha+a_\alpha^\dag), 
\end{equation}
while $K_{{\bf a},x}=J_x$ and $K_{{\bf a},y}= J_y$ for all the remaining links. We further require that 
\begin{equation}\label{multiple_readout_condition}
\m{if}~{\bf a}_{\alpha}\in\Omega,~\m{then}~{\bf a}_{\alpha}\pm {\bf e}_2 \notin\Omega,
\end{equation} 
such that two operators $W_{{\bf a}_{\alpha_1}}$, $W_{{\bf a}_{\alpha_2}}$ never share a $x$-link and thus no virtual process couples the $W_{{\bf a}_\alpha}$ operators to more than one cavity mode in lowest order. Then, the read-out of a single site can be generalized to obtain the following effective Hamiltonian:
\begin{equation}\label{eq:readout2}
H_{\rm eff}=\sum_{{\bf a}_{\alpha} \in \Omega} \left[\omega-\frac{2\delta_{x}^2}{4J_{z}-\omega} -\frac{4\delta_{x}^2J_{y}^2}{(4J_{z}-\omega)^3} W_{{\bf a}_{\alpha}}\right] a_\alpha^{\dagger}a_\alpha,
\end{equation}
Therefore, we can read-out simultaneously the eigenvalues of all $W_{{\bf a}_\alpha}$ through the shifts of the corresponding cavity mode frequencies. This procedure can be used to measure the state of the full memory in two steps only. This requires to have a cavity mode for each $W_{\bf a}$ operator and being able to turn on/off all the spin-electric couplings. Then, we can apply the above procedure twice in a checkerboard configuration where, in the first step, we turn on the couplings of only the plaquette operators and, in the second step, the couplings of only the star operators. 

\section{Resonant enhancement of the gap from a single cavity mode}\label{sec:resonant_enhancement}

We discuss in this and the following sections how the cavity modes can be used to prolong the lifetime of the encoded information. As shown in Sec.~\ref{sec:perturbativeapproach1}, the honeycomb model (\ref{eq:kitaevhoneycomb}) in the absence of cavity modes and with uniform couplings ($K_{{\bf a},x}=J_{x}$ and $K_{{\bf a},y}=J_{y}$) is equivalent to the toric code model of Eq.~(\ref{eq:toric}), up to fourth order in perturbation theory. This effective model can be used as a quantum memory encoding two qubits in the ground space, and the lifetime is essentially determined by the energy gap. \cite{KitaevToric} As it can be seen in Eq.~(\ref{eq:toric}), the excitations correspond to sites where $W_{\bf a} = -1$. They can also be interpreted as anyons, \cite{KitaevToric} with anyon numbers $n_{\bf a}$ defined as
\begin{equation}\label{ni_def}
W_{{\bf a}}=1-2n_{{\bf a}},
\end{equation}
at each site ${\bf a}$ of the lattice. If $W_{{\bf a}}=+1$, then no anyon is present, i.e., $n_{{\bf a}}=0$. Instead, if $W_{{\bf a}}=-1$ one anyon is present, i.e. $n_{{\bf a}}=+1$. In Eq.~(\ref{eq:toric}), the anyons are noninteracting particles with energy
\begin{equation}\label{mu_0}
\mu_0 = \frac{J_{x}^2J_{y}^2}{8J_{z}^3}.
\end{equation}
In the present section, we consider how to obtain a noninteracting toric model in the presence of a single cavity mode, and how the gap of the model is affected by the cavity. We refer to the next section for a discussion of anyon interactions induced by two cavity modes, as proposed in Ref. \onlinecite{ChesiPRA}. 

As for the single-site read-out discussed in Sec.~\ref{sec:readout}, the cavity mode has unperturbed frequency $\omega$ and annihilation operator $a$, but the coupling scheme is as follows, for all the $x$ links:
\begin{equation}\label{couplings_enhanced_gap}
K_{{\bf a},x}= J_x + \delta_x (a+a^\dag), 
\end{equation}
while $K_{{\bf a},y}= J_y$ for all the $y$ links. Our analysis can also be simply extended to non-homogeneous couplings, when $\delta_x$ becomes site dependent. The read-out scheme of Eq.~(\ref{couplings_single_readout}), where only two $x$ links are affected by the cavity, provides an explicit example.  

To obtain an effective Hamiltonian, we assume again here the resonant condition of Eq.~(\ref{resonance_cond}), which allows us to keep only the leading term (with resonant energy denominators) in the perturbative contributions of Eqs.~(\ref{SW_2}) and (\ref{SW_4}). The interested reader can find more details of the derivation in Appendix~\ref{app:SWONEMODE}. The final result reads
\begin{equation}\label{enhanced_gap}
H_{\rm eff}=\omega a^{\dagger}a-\frac{\delta_x^2N}{4J_{z}-\omega} a^{\dagger}a 
-\frac{4\delta_{x}^2J_{y}^2}{(4J_{z}-\omega)^3}a^{\dagger}a\sum_{{\bf a}}W_{{\bf a}},
\end{equation}
where the modified gap of the model is immediately seen to be
\begin{equation}\label{mu_resonant}
\mu =\frac{8\delta_{x}^2J_{y}^2}{(4J_{z}-\omega)^3}\langle a^\dag a\rangle,
\end{equation}
which is proportional to the number of photons populating the cavity mode. In fact, comparing $\mu$ to the bare gap $\mu_0$ of Eq.~(\ref{mu_0}), the quantity $\delta_x^2 \langle a^\dag a\rangle$ appears instead of $J_x^2$. Furthermore, the denominator is in terms of $(4J_{z}-\omega)$, which can be made smaller by simply changing the frequency of the cavity, in contrast to the bare value $J_z$. Indeed, we assumed $(4J_{z}-\omega) \ll J_{z}$, see Eq.~(\ref{resonance_cond}). These facts make the effect of that cavity very interesting, because the resonant gap $\mu$ can be made larger than $\mu_0$. Here and in the rest of this work, we assume that the cavity is driven out of equilibrium. The average $\langle a^\dag a\rangle$ appearing in Eq.~(\ref{mu_resonant}) corresponds to the nonequilibrium photon population of the cavity and can be large.

Note however that the gap cannot be made arbitrarily large. In particular, the perturbative expansion requires $\delta_x \sqrt{\langle a^\dag a\rangle},J_{x,y} < (4J_{z}-\omega)$, which implies $\mu < J_y$. Still, for given values $J_{x,y}\ll J_z$, an enhancement factor of the bare gap $\mu_0$ of order $J_{z}^3/J_{x}^2J_{y}$ can be achieved, by an appropriate design of the cavity and excitation of the resonant mode. Such increase in the gap has a dramatic effect on the lifetime of the quantum memory, if $\mu$ can become larger than the temperature $T$. In this case, the population of anyons is exponentially suppressed and the lifetime $\tau$ increases accordingly. The following approximate formula was proposed \cite{Alicki, Hamma2009,ChesiPRA}:
\begin{equation}\label{lifetime_formula}
\tau \simeq 2f_c \frac{e^{\mu/k_B T}+1}{D},
\end{equation}
where $D$ is the diffusion constant of the anyons and the prefactor $f_c$ can be interpreted as a critical fraction of errors. \cite{ChesiPRA} The value of $D$ depends on the details of the thermal bath, especially whether it is Ohmic or super-Ohmic, and can also contain a dependence on $\mu$ (see Ref. \onlinecite{ChesiPRA} for an extended discussion). 

Finally, we recall that the exact solution in the absence of cavity modes \cite{KitaevHoney} also displays a gapless phase, away from the perturbative regime (i.e., when $J_{x,y,z}$ have comparable values). The influence of the cavity mode on this gapless phase is an interesting question: Since in the resonant case the role of $4J_z$ is played by $4J_{z}-\omega$, it can be expected that not only the gap of the system can be modified by tuning $\omega$ around $4J_z$, but also a transition from the gapped to the gapless phase could be induced by changing the parameters of the cavity. These nonperturbative aspects will be the subject of future investigations.  

\section{LONG-RANGE INTERACTIONS}\label{sec:long-range}

It was shown in Ref. \onlinecite{ChesiPRA} that repulsive long-range interactions between anyons have a beneficial effect on the memory. Since cavity modes spatially uniform across the memory realize interactions with constant strength, we consider the following effective model:
\begin{equation}\label{interacting_anyons}
H_{\rm eff}= \mu \sum_{\bf a} n_{\bf a} + \frac{A}{2} \sum_{{\bf a}, {\bf a'}} n_{\bf a} n_{\bf a'},
\end{equation}
where ${\bf a},{\bf a'}$ run over all the $N$ sites of the lattice \cite{self_energy_comment} and the anyon numbers $n_{{\bf a}}=0,1$ are defined as in Eq.~(\ref{ni_def}). Note that, differently from Ref. \onlinecite{ChesiPRA}, we do not distinguish here between star and plaquette anyons, but all of them interact among each other. The effect of the interactions on the memory lifetime can be understood in terms of a mean-field gap $\mu_{\m{mf}}$, which includes the repulsion energy (if $A>0$) from the average anyon density $n_{\rm mf}$:
\begin{eqnarray}\label{eq:mf_gap}
\mu_{\m{mf}}&=&\mu +n_{\m{mf}} A N,
\end{eqnarray}
where $n_{\m{mf}}$ is determined self-consistently as $n_{\m{mf}}=1/(e^{\mu_{\m{mf}}/k_B T}+1)$ assuming an equilibrium density of anyons at temperature $T$. 

It is the purpose of this section to study how the model (\ref{interacting_anyons}) is realized perturbatively and when the effect of the interactions becomes relevant. A noticeable feature of Eq.~(\ref{eq:mf_gap}) is that, assuming constant $\mu,A$, the mean-field gap increases with $N$. This is because the contribution from the interaction (second term) grows and becomes eventually the dominant term for sufficiently large $N$. The effect is to reduce the anyon density $n_{\rm mf}$ and prolong the memory lifetime. \cite{ChesiPRA} However, two main differences appear in the explicit perturbative derivation:

(i) The parameters $\mu$ and $A$ acquire in general a nontrivial dependence on $N$. Therefore, it is possible that the anyon gap $\mu$ dominates the size dependence of the gap, instead of the interaction contribution.

(ii) The requirement to strictly remain in the perturbative regime imposes restrictions to the size of the system, below which in many cases the interaction contribution is small with respect to $\mu$. 

However, we also find specific coupling schemes and appropriate range of parameters for which the interactions become the dominant effect.

\subsection{Coupling scheme and resonant effective Hamiltonian}

As suggested in Ref. \onlinecite{ChesiPRA}, we consider two cavity modes with frequencies $\omega_{x,y}$ and annihilation operators $a_{x,y}$. To have two modes is useful because it allows to change the sign of the interaction, by choosing the frequencies and photon occupations of the two cavity modes. The first (second) mode couples to one-half of the $x(y)$-couplings. In particular, by referring to Fig.~\ref{fig:differentlinks}, we assume
\begin{equation}\label{half_coupling_x}
K_{{\bf a},x}=J_x + \delta_x(a_x^\dag +a_x), ~ \text{for the dark (red) $x$ links},
\end{equation}
while $K_{{\bf a},x}=J_x$ for the light (yellow) $x$ links and
\begin{equation}\label{half_coupling_y}
K_{{\bf a},y}=J_y + \delta_y(a_y^\dag +a_y), ~ \text{for the dark (green) $y$ links},
\end{equation}
while $K_{{\bf a},y}=J_y$ for the light (white) $y$ links.

We obtain the effective Hamiltonian for the toric code model from a SW transformation by assuming, as in the previous sections, that the two modes are resonant:
\begin{equation}
(4J_{z}-\omega_{x,y})  \ll J_{z},\omega_{x,y}.
\end{equation}
Although this considerably simplifies the treatment, still several contributions are present when evaluating the second and fourth order expressions, as discussed in more detail in Appendix  \ref{app:SW2cavities}. The following much simpler expression is obtained:
\begin{eqnarray}\label{eq:firstSWresonant}
&&H_{\m{eff}}=\omega'_x a_{x}^{\dagger}a_{x}+\omega'_y a_{y}^{\dagger}a_{y}\nonumber\\
&& -\frac12 \left[\sum_{\alpha=x,y} \frac{J_{x}J_{y}\delta_{x}\delta_{y}}{(4J_{z}-\omega_{\alpha})^3}\right]
\left(a_{x}^{\dagger}a_{y}+a_{x}a_{y}^{\dagger}\right)\sum_{{\bf a}}W_{{\bf a}},~~
\end{eqnarray}
by imposing the more restrictive condition 
\begin{equation}\label{condition_very_resonant}
|\omega'_x-\omega'_y|\ll  (4J_{z}-\omega_{x,y})  \ll J_{z},\omega_{x,y}.
\end{equation}
In Eqs.~(\ref{eq:firstSWresonant}) and (\ref{condition_very_resonant}) we defined
\begin{equation}\label{shifted_omega}
\omega'_\alpha = \omega_\alpha -\frac{N}{2}\frac{\delta_{\alpha}^2}{4J_{z}-\omega_{\alpha}},
\end{equation}
where the second term is generally a small correction to the frequencies, due to perturbative restrictions to the size of the system [see Eq.~(\ref{N_disconnected_bound})]. A feature of the scheme considered here is that no term diagonal in the photon modes which couples to $\sum_{\bf a} W_{\bf a}$  appears at fourth order. This is because each cavity mode only interacts with a single link of each $W_{\bf a}$ operator, and thus at most one photon operator per mode can appear in combination with $\sum_{\bf a} W_{\bf a}$. Therefore, all the couplings to $W_{\bf a}$ in the effective fourth-order Hamiltonian are off-diagonal in the photon operators. As a first approximation, due to Eq.~(\ref{condition_very_resonant}), we only kept one of these terms, namely the one appearing in the second line of Eq.~(\ref{eq:firstSWresonant}) which is resonant in the difference of the two cavity mode frequencies.

By performing a second SW transformation of Eq.~(\ref{eq:firstSWresonant}) in the photon modes (in the same way described in Ref. \onlinecite{ChesiPRA}) and keeping only terms involving spin operators, we obtain
\begin{equation}\label{eq:plaquettehamiltonian}
H_{\rm eff}=\frac{A}{8}\left(\sum_{{\bf a}}W_{{\bf a}}\right)^2,
\end{equation}
where $A$ is is given by
\begin{equation}\label{eq:A_formula_resonant}
A=2\left[\sum_{\alpha=x,y}
 \frac{J_x J_y\delta_x\delta_y}{(4J_{z}-\omega_{\alpha})^3}\right]^2\frac{\langle a^\dag_x a_x\rangle -\langle a^\dag_y a_y\rangle}{\omega'_{x}-\omega'_{y}}.
\end{equation}
By rewriting Eq.~(\ref{eq:plaquettehamiltonian}) in terms of anyons we obtain
\begin{equation}\label{eq:interactionWi}
H_{\rm eff}= -\frac{AN}{2} \sum_{\bf a} n_{\bf a} +\frac{A}{2}\sum_{{\bf a},{\bf a'}}n_{\bf a} n_{\bf a'}.
\end{equation}

As announced, the effective parameters $\mu$ and $A$ have a dependence on $N$ which in the case of the interaction strength $A$ can be quite weak [it appears through the denominator $\omega'_{x}-\omega'_{y}$, see Eqs.~(\ref{eq:A_formula_resonant}) and (\ref{shifted_omega})]. Instead, the chemical potential is approximately linear in $N$
\begin{equation}\label{mu_N}
\mu =-\frac{A N}{2},
\end{equation}
which makes it the dominant effect. Therefore, it is required for the stability of the $n_{\bf a}=0$ ground state that $A<0$, which leads to $\mu>0$ and can be realized by an appropriate choice of the frequencies $\omega'_{\alpha}$ and photon occupations $\langle a^\dag_\alpha a_\alpha\rangle$. The anyon interaction is in this case negative, which has the unfavorable effect of reducing the noninteracting gap, but it becomes quickly negligible with $N$. Note that, for a gap which grows linearly with system size, the lifetime is prolonged exponentially with $N$ at low temperature, as shown in Eq.~(\ref{lifetime_formula}).

\subsection{Interpretation in terms of anyon holes}

An interesting aspect of Eq.~(\ref{eq:plaquettehamiltonian}) is that it is symmetric with respect to the change of sign of all the $W_{\bf a}$. Therefore, it is useful to define the anyon-hole numbers
\begin{equation}
\bar{n}_{{\bf a}}=1-n_{{\bf a}},
\end{equation}
which describe excitations of the ground state with $W_{\bf a}=-1$ for every ${\bf a}$. Clearly, transforming Eq.~(\ref{eq:plaquettehamiltonian}) in terms of the $\bar{n}_{\bf a}$ leads to an Hamiltonian with the same form of Eq.~(\ref{eq:interactionWi}). Another interesting way to rewrite Eq.~(\ref{eq:plaquettehamiltonian}) in a symmetric way is by considering both types of particles present in the memory (i.e., at each site either an anyon or an anyon hole is present). This gives
\begin{equation}\label{eq:quasiparticlesinteractions}
H_{\rm eff}=-\frac{A}{2}\sum_{{\bf a},{\bf a'}}\bar{n}_{{\bf a}}n_{{\bf a'}},
\end{equation}
describing a long-range interaction between anyons and anyon holes. For $A<0$ such interaction is repulsive and the ground state is completely occupied with one type of particles, say anyon holes. The appearance of anyons in the memory then results in a repulsive contribution from the $\sim N$ anyon holes already present in the ground state. This picture provides a natural interpretation of the system size dependence of the gap, in terms of long-range interactions.

As a side remark, we consider again the model Hamiltonian Eq.~(\ref{interacting_anyons}), but with with constant coefficients $\mu,A$. By rewriting Eq.~(\ref{interacting_anyons}) in terms of anyon holes we obtain 
\begin{equation}\label{eq:attractiveHamiltonian}
H_{\rm eff}=-(\mu+A N)\sum_{\bf a} \bar{n}_{\bf a} +\frac{A}{2}\sum_{{\bf a},{\bf a'}}\bar{n}_{{\bf a}}\bar{n}_{{\bf a'}}.
\end{equation}
Differently from Ref.~\onlinecite{ChesiPRA}, we now assume $A<0$. Then, the first term of Hamiltonian (\ref{eq:attractiveHamiltonian}) is a size-dependent chemical potential which for large system size becomes positive, while the second term describes a long-range attractive interaction. Therefore, if we want to use this system as a quantum memory, we can encode a state in the memory full of anyons ($W_{\bf a}=-1$ for every ${\bf a}$, instead of $+1$). In this situation, the role of the anyons and the anyon holes is interchanged. For example, a spin-flip will now produce a pair of anyon holes which diffuse in the memory and destroy the information. However, since the chemical potential of the anyon holes increases linearly with the system size, one can prolong the lifetime of the memory by making the system size bigger.

\subsection{Off-resonant contributions to the anyon gap}\label{off_resonant_gap}

We have assumed so far the resonant condition (\ref{condition_very_resonant}), such that we could keep only a single term which couples to the star and plaquette operators [the second line of Eq.~(\ref{eq:firstSWresonant})]. However, many contributions to the gap and interactions exist in general and we show here that for some regime of parameters it is possible that $\mu$ in Eq.~(\ref{interacting_anyons}) becomes zero. In this case, the interactions between anyons have a dominant effect.

We consider again the effective photon Hamiltonian up to fourth order, obtained from Eqs.~(\ref{SW_1}) and (\ref{SW_4}), but keep two additional terms with respect to Eq.~(\ref{eq:firstSWresonant})
\begin{eqnarray}\label{eq:firstSWnotsoresonant}
H_{\m{eff}} && = \omega'_x a_{x}^{\dagger}a_{x}+\omega'_y a_{y}^{\dagger}a_{y}
-\frac{N}{4}\sum_{\alpha=x,y} \frac{J_{\alpha}\delta_{\alpha}}{4J_{z}-\omega_{\alpha}}\left(a_{\alpha}+a_{\alpha}^{\dagger}\right)\nonumber\\
&&-\left[\sum_{\alpha=x,y} \frac{J_{\bar{\alpha}}^2J_\alpha \delta_\alpha }{(4J_{z}-\omega_{\alpha})^3}(a_{\alpha}+a_{\alpha}^{\dagger})\right]\sum_{{\bf a}}W_{{\bf a}}\nonumber\\
&& -\left[\sum_{\alpha=x,y} \frac{J_{x}J_{y}\delta_{x}\delta_{y}}{2(4J_{z}-\omega_{\alpha})^3}\right]
\left(a_{x}^{\dagger}a_{y}+a_{x}a_{y}^{\dagger}\right)\sum_{{\bf a}}W_{{\bf a}},\end{eqnarray}
where in the second line $\bar\alpha=y$ if $\alpha=x$ and vice versa. For some details on the derivation of this expression, we refer to Appendix~\ref{app:SW2cavities}. In Eq.~(\ref{eq:firstSWnotsoresonant}), the second line generates now a chemical potential for the anyons in combination with the second-order term in the first line. Therefore, Eq.~(\ref{mu_N}) gets modified as follows:
\begin{eqnarray}\label{eq:chemicalpotentialvanishes}
\mu=-\frac{AN}{2}+\sum_{\alpha=x,y}\frac{J_x^2J_y^2 \delta_{\alpha}^2}{(4J_{z}-\omega_{\alpha})^4 \omega^\prime_\alpha}N.
\end{eqnarray}
Equation~(\ref{eq:chemicalpotentialvanishes}) shows that the effect of the off-resonant terms becomes negligible if $\omega_{\alpha}^{'}$ is sufficiently large. However, this new contribution to the chemical potential appears to lower order in the perturbation expansion than the first term (sixth order in the parameters $J_{x,y},\delta_{x,y}$ instead of eighth order). Therefore, it is possible to make the off-resonant contribution of the same size of the resonant one by slightly relaxing the condition (\ref{condition_very_resonant}). Note that the second term in Eq.~(\ref{eq:chemicalpotentialvanishes}) is always positive and can only be canceled if $A>0$, which is possible by choosing suitable frequencies and photon occupations of the two modes [see Eq.~(\ref{eq:A_formula_resonant})]. The anyon interaction is therefore repulsive and the final Hamiltonian is similar to the case studied in Ref. \onlinecite{ChesiPRA}.

\begin{figure}
	\centering
		\includegraphics[width=0.45\textwidth]{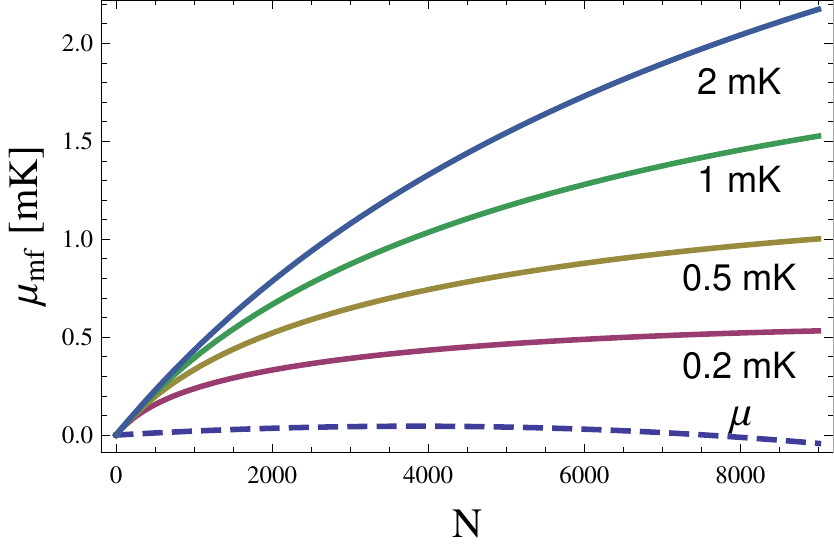}
	\caption{Mean-field gap $\mu_{\m{mf}}$, given by Eq. (\ref{eq:mf_gap}) with $A$ as in Eq.~(\ref{eq:A_formula_resonant}), plotted for different temperatures (solid curves). The dashed curve is the bare gap $\mu$ of Eq.~(\ref{eq:chemicalpotentialvanishes}). We used $J_{z}=1011$ K, $J_{x_{1,2}}=J_{y_{1,2}}=22$ K, $\langle a_{x}^{\dagger}a_{x}\rangle=100$, $\langle a_{y}^{\dagger}a_{y}\rangle=0$, $\delta_{x}=\delta_{y}=0.21$ K, $\omega_{x}=3991$ K, and $\omega_{y}=3980$ K.}
	\label{fig:MeanFieldGap}
\end{figure}

As a specific example, we plot $\mu$ in Fig.~\ref{fig:MeanFieldGap} as function of $N$, by using in Eq.~(\ref{eq:chemicalpotentialvanishes}) suitable numerical values of the couplings, such that $\mu$ is approximately zero. Since the dependence on $N$ of Eq.~(\ref{eq:chemicalpotentialvanishes}) is not exactly linear for both terms (due to $A$ and $\omega^\prime_\alpha$), $\mu$ cannot be made zero identically for arbitrary $N$. Nevertheless, in a range of parameters, it is sufficiently small that the repulsive anyon interaction becomes the dominant effect. This we show by plotting in Fig.~\ref{fig:MeanFieldGap} the mean-field gap $\mu_\m{mf}$, obtained from Eq.~(\ref{eq:mf_gap}), which can be much larger than the noninteracting gap $\mu$.

The effect of the interactions is very beneficial since it allows to obtain $\mu_{\m{mf}}$ in the mK range, while the original gap without cavities is $J_{x}^2J_{y}^2/8J_{z}^3\simeq 30~\mu$K. However, although the interacting regime presents interesting aspects, it seems more useful in practice to take advantage of the resonant enhancement of the gap discussed in Sec.~\ref{sec:resonant_enhancement}. This is realized if each mode couples to all the corresponding ($x$- or $y$-) links, instead of half of them. The final result is the same of Eq.~(\ref{mu_resonant}), including two resonant modes
\begin{equation}\label{mu_resonant_2}
\mu \simeq \sum_\alpha \frac{8\delta_{\alpha }^2J_{\bar\alpha}^2}{(4J_{z}-\omega_\alpha)^3}\langle a_\alpha^\dag a_\alpha\rangle.
\end{equation}
With the same parameters of Fig.~\ref{fig:MeanFieldGap}, Eq.~(\ref{mu_resonant_2}) gives $\mu \simeq 0.1$~K, much larger than the maximum value $\mu_{\rm mf}\simeq 2$ mK of Fig.~\ref{fig:MeanFieldGap}. As discussed, the effect of the interaction becomes larger with system size, but unfortunately the perturbative treatment is only strictly applicable within a limited range of values of $N$. Conditions for $N$ are discussed in Appendixes~\ref{app:disconnected_terms} and \ref{app:SW2cavities}, and lead to an upper bound for $N$ similar to the largest value in Fig.~\ref{fig:MeanFieldGap}.

\section{CAVITY MODES OUT OF RESONANCE}\label{sec_not_resonant}

In the previous sections, we generally assumed that the different cavity modes are resonantly coupled to the memory, i.e., $\omega_\alpha \simeq 4 J_z$. We now consider the case of largely detuned modes. Since cavity modes with large frequency $\omega_\alpha \gg J_z$ have a small effect on the spin model, we consider the opposite limit:
\begin{equation}\label{small_frequency}
\omega_{\alpha}\ll4J_{z},
\end{equation}
and show that many results of the previous sections have an analog in this regime. As before, we can still compute the SW effective Hamiltonian given by Eqs. (\ref{SW_2}) and  (\ref{SW_4}), but it is not possible now to pick a few resonant terms at each order of perturbation theory. However, as shown in Appendix~\ref{app:offresonant}, the final result can be obtained from the calculation without cavity modes, by substituting $J_{{\bf a},x}$, $J_{{\bf a},y}$ with the corresponding full couplings (\ref{eq:couplings}): 
\begin{eqnarray}\label{prescription_non_res}
J_{{\bf a},x} \to K_{{\bf a},x},\qquad 
J_{{\bf a},y} \to K_{{\bf a},y}.
\end{eqnarray}
Given this simple prescription, the expression Eq.~(\ref{eq:SWTtwocouplings}) for the Kitaev model with four distinct couplings $J_{x_{1,2}}$, $J_{y_{1,2}}$ proves to be useful for several configurations.

We consider first a single cavity mode coupled homogeneously to the $x$ links as in Eq.~(\ref{couplings_enhanced_gap}), which amounts to set 
\begin{equation}
J_{x_{1,2}} \to J_x+\delta_x(a+a^\dag), \quad J_{y_{1,2}}\to J_y,
\end{equation}
in Eq.~(\ref{eq:SWTtwocouplings}). As in Sec.~\ref{sec:resonant_enhancement}, we can neglect the off-diagonal terms in the photon Hamiltonian and obtain
\begin{equation}\label{enhanced_gap_offres}
H_{\rm eff}= \left(\omega -\frac{\delta_x^2 N}{2J_{z}}\right) a^{\dagger}a 
-\frac{J_{y}^2[J_x^2+\delta_{x}^2(2a^\dag a+1)]}{16J_z^3}\sum_{{\bf a}}W_{{\bf a}},
\end{equation}
which shows an enhancement of the gap from the cavity mode. Even if the gap from Eq.~(\ref{enhanced_gap_offres}) does not have a resonant denominator, it is still possible to increase it by driving the cavity to a large photon population. From the point of view of the perturbative treatment, the only condition to be satisfied in Eq.~(\ref{enhanced_gap_offres}) for the number of photons is $\delta_x \sqrt{\langle a^\dag a \rangle } \ll J_z$.

We now consider the anyon interactions and choose the same coupling scheme of Sec.~\ref{sec:long-range} with two cavity modes $a_{x,y}$ respectively coupled to half of the $x$- or $y$ links. More explicitly, we set in Eq.~(\ref{eq:SWTtwocouplings})
\begin{equation}\label{interacting_scheme_offres}
J_{\alpha_{1}} \to J_\alpha+\delta_\alpha(a_\alpha+a^\dag_\alpha), \quad J_{\alpha_2}\to J_\alpha,
\end{equation}
where $\alpha=x,y$. By keeping only terms which are relevant for the star and plaquette operators (see Appendix~\ref{app:offresonant}), we obtain
\begin{eqnarray}\label{Heff_nonres_2cavities}
H_{\m{eff}}&=&\omega^{\prime\prime}_{x}a_{x}^{\dagger}a_{x}+\omega^{\prime\prime}_{y}a_{y}^{\dagger}a_{y}
-\frac{N}{4J_{z}}\sum_{\alpha=x,y} J_{\alpha}\delta_{\alpha}(a_{\alpha}+a_{\alpha}^{\dagger})\nonumber\\
&-&\frac{J_{x}^2J_{y}^2}{16J_{z}^3}\sum_{{\bf a}}W_{{\bf a}}-\left[\sum_{\alpha=x,y}\frac{J_{\bar\alpha}^2 J_{\alpha}\delta_{\alpha}}{16J_{z}^3}(a_{\alpha}+a_{\alpha}^{\dagger})\right]\sum_{{\bf a}}W_{{\bf a}}\nonumber\\
&-&\frac{J_{x}J_{y}\delta_{x}\delta_{y}}{16J_{z}^3}\left(a_{x}a_{y}^{\dagger}+a_{y}^{\dagger}a_{x}\right)\sum_{{\bf a}}W_{{\bf a}}
\end{eqnarray}
which is in complete analogy with Eq.~(\ref{eq:firstSWnotsoresonant}), except that the first $\sum_{\bf a} W_{\bf a}$ term in the second line (present even without cavity modes) is not neglected here, since all the energy denominators are non-resonant. In Eq.~(\ref{Heff_nonres_2cavities}) we defined
\begin{equation}\label{omega_double_prime}
\omega^{\prime\prime}_\alpha=\omega_\alpha-\frac{N}{4J_{z}}\delta_{\alpha}^2,
\end{equation}
where the second term is generally a small correction in the perturbative regime. As in the previous section, we perform a second SW transformation in the photon field and obtain the following effective Hamiltonian by making use of the resonant condition $\omega_{x}^{\prime\prime}\simeq \omega_{y}^{\prime\prime}$
\begin{equation}\label{eq:offresonance}
H=\frac{A}{8}\left(\sum_{{\bf a}}W_{{\bf a}}\right)^2,
\end{equation}
where
\begin{equation}\label{eq:B}
A=2 \left(\frac{J_{x}J_{y}\delta_{x}\delta_{y}}{8J_{z}^3}\right)^2
\frac{\langle a^\dag_x a_x\rangle-\langle a^\dag_y a_y\rangle}{\omega^{\prime\prime}_{x}-\omega^{\prime\prime}_{y}}.
\end{equation}
Similar to Sec.~\ref{sec:long-range}, we can rewrite Eq.~(\ref{eq:offresonance}) in terms of the anyon numbers $n_{{\bf a}}$, in the form of Eq.~(\ref{interacting_anyons}), with $A$ given by Eq.~(\ref{eq:B}) above and a chemical potential $\mu=-AN/2$, approximately proportional to the system size. We recall again that $A$ can be made either positive or negative, depending on the sign of $\langle a^\dag_x a_x\rangle-\langle a^\dag_y a_y\rangle$ and $(\omega^{\prime\prime}_{x}-\omega^{\prime\prime}_{y})$, with $A<0$ leading to a stable memory. As in Sec.~\ref{sec:long-range}, if we stay a little bit away from the $\omega_{x}^{\prime\prime}\simeq \omega_{y}^{\prime\prime}$ resonance, the lower-order terms become relevant and we can write the following expression for the chemical potential of the anyons: 
\begin{eqnarray}
\mu = -\frac{AN}{2}+\frac{J_{x}^2J_{y}^2}{8J_{z}^3}
+\sum_{\alpha=x,y}\frac{J_{x}^2J_{y}^2\delta_{\alpha}^2 N}{16J_{z}^4 \omega^{\prime\prime}_\alpha},
\end{eqnarray}
which can be put to zero by choosing $A>0$  and for some specific choice of parameters. This allows to recover the same situation of 
Sec.~\ref{sec:long-range}, when the dynamics of the system is solely determined by the long-range positive interaction between the anyons. 

\section{Conclusion}

In this paper we have analyzed a generalization of the Kitaev honeycomb model, which includes the influence of quantized cavity modes. The coupling is realized by a linear modification of the Ising interactions which should be relevant for a variety of systems. \cite{MirceaPRL, MirceaPRB, CanaliPRB, Kugel, MirceaCoulomb, PashkinNature, AverinBruder, Nori, KitaevMott, MuellerNaturePhysics, LukinPRL,Blais2004,WallraffPRL,Angelakis} We have studied this model in the perturbative regime, when the low-energy Hamiltonian is a generalization of the toric code and the setup can thus be seen as a quantum memory. We have considered different coupling schemes and shown that the presence of cavity modes can be useful in several ways.

More specifically, cavity modes allow to read out the state of the star and plaquette operators. An efficient read-out scheme of the error syndrome is an essential prerequisite to perform any error-correction algorithm. \cite{Dennis,Poulin} The simplest way to achieve this goal is to couple one cavity mode to the desired star or plaquette and to detect the corresponding frequency shift, similarly to a read-out scheme for single superconducting qubit in circuit quantum electrodynamics. \cite{Blais2004,WallraffPRL} An array of multiple resonators allows to measure all the star or plaquette operators simultaneously.

Secondly, we studied cavity modes coupled homogeneously across the whole memory. We have considered both setups with one and two cavity modes, resonantly coupled or with very small frequencies. For the case of a single cavity we have obtained the dependence of the anyon gap from the photon number and frequency of the mode, and shown that a large enhancement is possible with an excited cavity at resonance. Having two cavity modes allows to realize long-range interactions with tunable sign among anyons and a size-dependent chemical potential. In this case, the relevant excitation energy is a mean-field gap \cite{ChesiPRA} which, in some parameter regimes, is essentially due to the anyon interactions and is generally much larger than the original gap without cavities. Since the lifetime of the memory depends exponentially on the value of the gap at low temperature, these results might lead to a dramatic prolongation of the memory lifetime. From a more general perspective, the extension of the honeycomb Hamiltonian considered in this work allows to have additional control on the properties of the model, through the specific features of the coupling scheme and the possibility of tuning the parameters characterizing the cavity modes.

\section{Acknowledgments}
We would like to thank Mircea Trif for insightful suggestions. We also acknowledge useful discussions with Jan Fischer, Dimitrije Stepanenko, and Ying-Dan Wang. This work was supported by the Swiss
NSF, NCCR Nanoscience Basel, the DARPA QUEST program, and the EU project SOLID.

\appendix

\section{SW TRANSFORMATION}\label{appendix_SW}

For the convenience of the reader, we summarize here the formalism of the Schrieffer-Wolff (SW) transformation \cite{Phys.Rev.149.491,SWunpublished} and show how it can be applied to the general class of Hamiltonians of interest in this work.

\subsection{General formalism}

Consider a Hamiltonian $H$ with a projector $P_H$ on an invariant subspace (i.e., a direct sum of eigenspaces) and two known projectors $P$ and $Q=1-P$, such that the subspace of $P$ has the same dimension of the one of $P_H$. A SW transformation is defined as a unitary transformation $U=e^{S}$ with block off-diagonal $S$ which maps the subspace of $P_H$ into the one of $P$. Therefore, $P=e^S P_H$ and the transformed Hamiltonian $H_{\m{d}}=UHU^{\dagger}$ is block diagonal:
\begin{equation}
PSP=QSQ=0,
\end{equation}
and
\begin{equation}
PH_{\m{d}}Q=QH_{\m{d}}P=0.
\end{equation}
It is useful to define the superoperator $L$ as
\begin{equation}
LA=[S,A],
\end{equation}
such that, in the superoperator language, the transformed Hamiltonian $H_{\m{d}}$ takes the following compact form: 
\begin{equation}
H_{\m{d}}=e^{L}H.
\end{equation}

Let us now consider an Hamiltonian that can be decomposed in a dominant part $H_{0}$ and a small perturbation $V=V_{\m{d}}+V_{\m{od}}$, where the spectrum of $H_{0}$ is divided in a low-energy space ($P$ projects onto the low-energy space) and a high-energy space ($Q=1-P$ projects onto the high-energy space). By definition, the diagonal perturbation $V_{\m{d}}$ and the off-diagonal perturbation $V_{\m{od}}$ satisfy the following equations
\begin{eqnarray}
PV_{\m{d}}Q&=&QV_{\m{d}}P=0,\nonumber\\
PV_{\m{od}}P&=&QV_{\m{od}}Q=0.
\end{eqnarray}
Since $V$ is assumed to be small, we can expand $S$ and $L$ in a series, namely
\begin{equation}
S=\sum_{n=1}S_{n},\quad {\rm and} \quad L=\sum_{n=1}L_{n},
\end{equation}
where $L_{n}A:=\left[S_{n},A\right]$. If we require that the off-diagonal part vanishes at each order, we find the following recursive relations for the $S_{n}$ operators\cite{SWunpublished} ($n=1,2,3,...$):
\begin{eqnarray}
&&S_{1}=L_{0}^{-1}V_{\m{od}},\label{eq:s1}\\
&&S_{2}=L_{0}^{-1}L_{1}V_{\m{d}},\label{eq:s2}\\
&&S_{3}=-\frac{1}{3}L_{0}^{-1}L_{1}^{3}H_{0}+L_{0}^{-1}L_{2}V_{\m{d}}\label{eq:s3}, \\
&&\ldots, \nonumber
\end{eqnarray}
where $L_{0}A:=\left[H_{0},A\right]$. It is now possible to find general expressions for the low-energy effective Hamiltonian at each order, namely
\begin{eqnarray}
&&H_{\m{eff}}^{(1)}=PV_{\m{d}}P,\label{eq:h1}\\
&&H_{\m{eff}}^{(2)}=\frac{1}{2}PL_{1}V_{\m{od}}P,\label{eq:h2}\\
&&H_{\m{eff}}^{(3)}=\frac{1}{2}PL_{2}V_{\m{od}}P,\label{eq:h3}\\
&&H_{\m{eff}}^{(4)}=\frac{1}{2}PL_{3}V_{\m{od}}P-\frac{1}{24}PL_{1}^3V_{\m{od}}P,\label{eq:h4}\\
&&\ldots. \nonumber
\end{eqnarray}

\subsection{Application to a general class of Hamiltonians}\label{app:generalclass}

In this section, we want to apply the formalism presented above to a general class of Hamiltonians relevant to this work:
\begin{eqnarray}\label{eq:generalclass}
H&=& H_{0}+\sum_{j,n} T_{n}^{j},
\end{eqnarray}
defined on a product Hilbert space $\mathcal{H}_{1}\otimes\mathcal{H}_{2}$. From Eq.~(\ref{Tn_sum}), it is clear that $H$ of Eq.~(\ref{eq:kitaevhoneycomb}) has the above form. The unperturbed spectrum (given by $H_0$) can be described in terms of a single type of excitations in $\mathcal{H}_1$ (in our case, hard-core bosons) and a set of excitations in $\mathcal{H}_2$, labeled by an index $\alpha$ (in our case, photons belonging to different cavity modes). The operators $T_{n}^{j}$ acts on $\mathcal{H}_{1}$ by changing the number of hard-core bosons by $n$ (in our case, $n=0,\pm2$) and on $\mathcal{H}_{2}$ according to the index $j$. This can assume the values $j=m\alpha$, indicating that the number of photons of type $\alpha$ changes by $m$ (in our case, $m=0,\pm 1$). As discussed in the main text, we also use the convention $0\alpha \equiv 0$, such that there is a unique set of $T^0_n$ operators (which do not change the photon number of any mode).

At this point, we define the low-energy space of this system as the lowest energy eigenspace in $\mathcal{H}_{1}$. Therefore, $P$ is the projector onto the subspace with no hard-core bosons, and $Q=1-P$ onto the subspace where some hard-core boson is present. The diagonal and off-diagonal perturbations are given by
\begin{eqnarray}
&& V_{\m{d}}=\sum_{j}T_{0}^{j}+Q\underbrace{\sum_{j} \sum_{n\neq 0} T_{n}^{j}}_{=:B}Q,\label{eq:diago}\\
&& V_{\m{od}}=PBQ+QBP.
\end{eqnarray}

We assume in the following, as appropriate for our case,
\begin{equation}\label{T0_P_0}
P T_0^j = T_0^j P =0,
\end{equation}
and that the energy change due to the $T_{n}^{j}$ operators is given by $\delta\epsilon_{n}^{j}$. Therefore, the following commutation relations hold:
\begin{equation}\label{eq:commutation2}
\left[H_{0},T_{n}^{j}\right]=\delta\epsilon_{n}^{j} \, T_{n}^{j}.
\end{equation}
In the simplest case, the excitations also correspond to energy quanta, as in Eq.~(\ref{commutations2}).

It is now possible to calculate $S_{1,2,3}$ according to Eqs.~(\ref{eq:s1}-\ref{eq:s3}), which give
\begin{eqnarray}
S_{1}&=&\sum_{i}{\sum_n} \left[ \frac{PT_{n}^{i}Q}{\delta\epsilon_{n}^{i}}-{\rm H.c.} \right],\label{eq:ss1}\\
S_{2}&=&\sum_{i,j}{\sum_{n,m}} \left[ \frac{PT_{n}^{i}QT_{m}^{j}Q}{\delta\epsilon_{n}^{i}(\delta\epsilon_{n}^{i}+\delta\epsilon_{m}^{j})}-{\rm H.c.}\right],\label{eq:ss2}\\
S_{3}&=&\sum_{i,j,k}{\sum_{n,m,l}}  \left[ \frac{PT_{n}^{i}QT_{m}^{j}QT_{l}^{k}Q}{\delta\epsilon_{n}^{i}(\delta\epsilon_{n}^i+\delta\epsilon_{m}^{j})(\delta\epsilon_{n}^{i}+\delta\epsilon_{m}^{j}+\delta\epsilon_{l}^{k})}\right. \nonumber\\
&&+\frac13 
\frac{ P T_{n}^{i} Q T_{l}^{k}PT_{m}^{j}Q+PT_{m}^{j} Q T_{n}^{i} P T_{l}^{k}Q}{\delta\epsilon_{n}^{i}\delta\epsilon_{l}^{k}(\delta\epsilon_{n}^{i}+\delta\epsilon_{m}^{j}+\delta\epsilon_{l}^{k})}
\nonumber \\
&&   \left. - \frac23 \frac{ P T_{n}^{i} Q T_{m}^{j}PT_{l}^{k}Q}{\delta\epsilon_{n}^{i}\delta\epsilon_{l}^{k}(\delta\epsilon_{n}^{i}+\delta\epsilon_{m}^{j}+\delta\epsilon_{l}^{k})}   
-{\rm H.c.}\right]\label{eq:ss3},
\end{eqnarray}
where every term with $n=0$ vanishes because of Eq.~(\ref{T0_P_0}) and H.c. denotes Hermitian conjugates (note that all the $S_p$ are anti-Hermitian). These Hermitian conjugate terms can be written in a form similar to the ones explicitly appearing in the square parenthesis, by making use of
\begin{equation}
\left( T_{n}^i\right)^\dag=T_{-n}^{-i},\quad {\rm and} \quad \delta\epsilon_{-n}^{-i}=-\delta\epsilon_{n}^{i}.
\end{equation}

With the help of Eqs.~(\ref{eq:h1}-\ref{eq:h4}), we finally obtain the low-energy effective Hamiltonian up to fourth order:
\begin{eqnarray}
&&\label{SW_1}H_{\m{eff}}^{(1)}= 0,\\
&&\label{SW_2}H_{\m{eff}}^{(2)}= \frac{1}{2}\sum_{i,j}{\sum_{n,m}} \left[ \frac{PT_{n}^{i}QT_{m}^{j}P}{\delta\epsilon_{n}^{i}}+ {\rm H.c.}\right],\\
&&\label{SW_3}H_{\m{eff}}^{(3)}=\frac{1}{2}\sum_{i,j,k}{\sum_{n,m,l}}
\left[\frac{PT_{n}^{i}QT_{m}^{j}QT_{l}^{k}P}{\delta\epsilon_{n}^{i}(\delta\epsilon_{n}^{i}+\delta\epsilon_{m}^{j})}+{\rm H.c.}\right], \\
&&\label{SW_4}H_{\m{eff}}^{(4)}=\frac12 \sum_{i,j,k,r}{\sum_{n,m,l,q}}\hspace{0.2cm}
\left[
\frac23 \frac{-PT_{n}^{i}QT_{m}^{j}PT_{l}^{k}QT_{q}^{r}P}{\delta\epsilon_{n}^{i}\delta\epsilon_{l}^{k}(\delta\epsilon_{n}^{i}+\delta\epsilon_{m}^{j}+\delta\epsilon_{l}^{k})}
\right.\nonumber\\
&& +\frac13 \frac{PT_{n}^{i}QT_{l}^{k}PT_{m}^{j}QT_{q}^{r}P+PT_{m}^{j}QT_{n}^{i}PT_{l}^{k}QT_{q}^{r}P}{\delta\epsilon_{n}^{i}\delta\epsilon_{l}^{k}(\delta\epsilon_{n}^{i}+\delta\epsilon_{m}^{j}+\delta\epsilon_{l}^{k})}\nonumber \\
&& -\frac{1}{12} \frac{PT_{n}^{i}QT_{m}^{j}PT_{l}^{k}QT_{q}^{r}P+3PT_{n}^{i}QT_{q}^{r}PT_{m}^{j}QT_{l}^{k}P}{\delta\epsilon_{n}^{i}\delta\epsilon_{m}^{j}\delta\epsilon_{l}^{k}}\nonumber\\
&&\left.  +\frac{PT_{n}^{i}QT_{m}^{j}QT_{l}^{k}QT_{q}^{r}P}
{\delta\epsilon_{n}^{i}(\delta\epsilon_{n}^{i}+\delta\epsilon_{m}^{j})(\delta\epsilon_{n}^{i}+\delta\epsilon_{m}^{j}+\delta\epsilon_{l}^{k})} +{\rm H.c.} \right],
\end{eqnarray}
where again H.c. are Hermitian conjugate terms.

\section{SW TRANSFORMATION WITH A SINGLE  RESONANT CAVITY MODE}\label{app:SWONEMODE}

We consider in this appendix the application of the SW formalism to the case when a single cavity mode is present, coupled only to $x$ links. This is useful for  both Secs.~\ref{sec:readout} and \ref{sec:resonant_enhancement}, where a read-out scheme and the effect of the cavity mode on the gap were studied.

\subsection{Derivation of the effective Hamiltonian}

In this case, the perturbation is given by $T_n^0$ and $T_n^{\pm}$ operators, where we drop the index $\alpha$ since only one mode is present. The treatment of the previous Appendix \ref{app:generalclass} can be easily applied to this case, where $\delta\epsilon_{n}^{j}$ of Eq.~(\ref{eq:commutation2}) is as in Eq.~(\ref{commutations2}), with $\omega_\alpha=\omega$. Since we are interested in the resonant case $(4J_{z}-\omega) \ll 4J_{z},\omega$, the relevant energy differences are
\begin{eqnarray}
&&\delta\epsilon_{+2}^-=-\delta\epsilon_{-2}^+=4J_z-\omega, \\
&&\delta\epsilon_0^0=0.
\end{eqnarray}
It is then possible to directly apply Eqs.~(\ref{SW_2}) and (\ref{SW_4}), keeping only the terms where the energy denominators are obtained from $\delta\epsilon_{+ 2}^-$, $\delta\epsilon_{- 2}^+$, and $\delta\epsilon_0^0$, which leads to
\begin{eqnarray}\label{Heff_resonant}
&&H_{\m{eff}}=\omega a^{\dagger}a-\frac{1}{4J_{z}-\omega}\left( \frac12 T_{-2}^{+}T_{+2}+\m{H.c.} \right)\nonumber\\
&&+ \frac{1}{(4J_{z}-\omega)^3}\left(\frac{5}{8}T_{-2}^{+}T_{+2}^{-}T_{-2}^{+}T_{+2}-\frac{1}{4} T_{-2}^{+}T_{-2}^{+}T_{+2}^{-}T_{+2}  \right. \nonumber\\
&&\left. -\frac{1}{8}T_{-2}^{+}T_{+2}T_{-2}^{+}T_{+2}^{-}-\frac{1}{2} T_{-2}^{+}T_{0}^{0}T_{0}^{0}T_{+2}+\m{H.c.}\right),
\end{eqnarray}
where $T_{+2}=T^0_{+2}+T^{+}_{+2}+T^{-}_{+2}$, as defined in Eq.~(\ref{Tn_sum}). 

The effective Hamiltonian (\ref{Heff_resonant}) is valid in the subspace with zero hard-core bosons, but is not diagonal in the number of photons. Therefore, it represents an effective Hamiltonian for the cavity and it requires to be further diagonalized to obtain the energy eigenstates. The explicit form of the photon Hamiltonian depends on the specific coupling scheme of the cavity to the links of the model, which is reflected in the $T_n^{j}$ operators. As an example, we assume the homogeneous coupling scheme (\ref{couplings_enhanced_gap}), to all the $x$ links of the model, and consider first the terms up to second order. Evaluating the first line of Eq.~(\ref{Heff_resonant}) in terms of effective spins and photon operators gives 
\begin{eqnarray}\label{eq:perturbationexpansion}
H_{\m{eff}}=&&~\omega a^{\dagger}a-\frac{\delta_{x}^2N}{4J_{z}-\omega}a^{\dagger}a
-\frac{J_x\delta_x}{4J_{z}-\omega}\left(a+a^{\dagger}\right)\frac{N}{2} \nonumber\\
&& -\frac{\delta_x^2}{4J_{z}-\omega}\left(a^2+a^{\dagger2}\right)\frac{N}{2}+\ldots .
\end{eqnarray}
The first two terms appear in Eq.~(\ref{enhanced_gap}) while the off-diagonal terms can be eliminated with a second SW transformation, in the photon operators only. Such off-diagonal terms involve excitations with energy $\omega$ or $2\omega$ and therefore give small corrections in the resonant limit we are interested. Therefore, we have neglected them in Eq.~(\ref{enhanced_gap}).

Evaluating all the fourth-order terms appearing in Eq.~(\ref{Heff_resonant}) is cumbersome, but this is not necessary for our purposes. As explained above, we keep only the terms already diagonal in the photon number, while the other ones give only small nonresonant corrections. Therefore, we can simply substitute all the $T_n$ operators in Eq.~(\ref{Heff_resonant}) by $T^-_n$. Furthermore, we are interested to the coupling of the photon mode to the $W_{\bf a}$ operators, which are the product of spin operators involving four \emph{distinct} links of the model (of which two are $x$ links and two $y$ links). To understand the relevant terms at fourth order we note that, for the specific coupling scheme (\ref{couplings_enhanced_gap}), only the $x$ links are coupled to the cavity mode. Therefore: (i) the $T^0_n$ operators involve a sum of terms with all the links of the model, while (ii) the $T^\pm_n$ operators are sums of terms with only $x$ links. Since only the $T_{0}^{0}$ operators contain terms relative to the $y$ links, it becomes clear that only the $T_{-2}^{+}T_{0}^{0}T_{0}^{0}T^-_{+2}$ combination can couple to the $W_{\bf a}$ operators. Finally, (\ref{Heff_resonant}) can be simplified to the following form
\begin{equation}\label{Heff_resonant_simple}
H_{\m{eff}}=\omega a^{\dagger}a-\frac{T_{-2}^{+}T^-_{+2}}{4J_{z}-\omega}- \frac{T_{-2}^{+}T_{0}^{0}T_{0}^{0}T^-_{+2}}{(4J_{z}-\omega)^3},
\end{equation}
which evaluates to Eq.~(\ref{enhanced_gap}), by expressing it in terms of photon and effective spin operators. As it turns out, the same reasonings apply to the read-out methods we discussed in Sec.~\ref{sec:readout}, where as well only couplings to the $x$ links appear. Therefore, Eq.~(\ref{Heff_resonant_simple}) is also the relevant effective Hamiltonian for the coupling schemes (\ref{couplings_single_readout}) and (\ref{couplings_multiple_readout}), which evaluate to Eqs.~(\ref{eq:readout}) and (\ref{eq:readout2}) respectively. Note that we have always neglected in the final results a subleading constant correction of the cavity frequency, i.e., the $a^\dag a$ term appearing in fourth order and independent of the $W_{\bf a}$ operators.

\subsection{Disconnected terms}\label{app:disconnected_terms}

\begin{figure}
	\centering
		\includegraphics[width=0.30\textwidth]{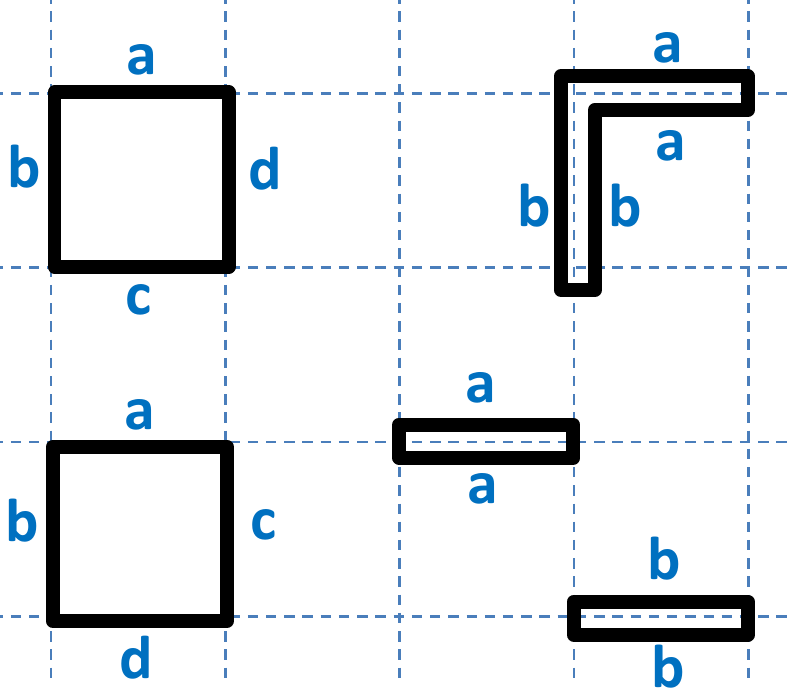}
	\caption{Pictorial representation of the connected and disconnected terms entering the perturbation expansion (\ref{Heff_resonant}). Both loops $abcd$ and $abba$ represent the different contributions arising from the connected term $T_{-2}^{+}T_{0}^{0}T_{0}^{0}T_{+2}^{-}$. We see that only $abcd$ loops lead to terms proportional to $W_{\bf a}$, since all four links are different. This is not the case for the $abba$ loop and the corresponding terms are constants. The remaining fourth-order terms entering Eq.~(\ref{Heff_resonant}) can be represented by two disconnected loops $aa$ and $bb$. Since the number of different positions of each $aa$ and $bb$ loop is proportional to $N$, the contribution of the disconnected terms is $\sim N^2$.}
	\label{fig:FeynmanDiagrams2}
\end{figure}

We would like to comment here on a special aspect which arises in the presence of cavity modes, i.e., the appearance in the perturbative expansion of non-extensive terms in the size of the system $N$. These originate from disconnected terms in the perturbation expansion which, if the Hamiltonian is local, give a vanishing contribution. \cite{SWunpublished} This result, known as linked-cluster theorem, is not valid here because of the long-range nature of the coupling of the photon modes, which are delocalized and can interact simultaneously with links at any distance across the lattice. 

For the case of interest here, the only connected fourth-order term appearing in Eq.~(\ref{Heff_resonant}) is the last one, with its Hermitian conjugate. Consider in fact the action of $T_{-2}^{+}T_{0}^{0}T_{0}^{0}T_{+2}$: the first operator from right to left, $T_{+2}$, creates two hard-core bosons on a certain link. The second and third operators make such hard-core bosons hop, and therefore they have to act on links where only one hard-core boson is present. Finally, the last operator $T_{-2}^{+}$ annihilates the two hard-core bosons. Therefore the links involved form a single closed loop, of the form $abcd$ (if all four links are different, which contributes with a term proportional to $W_{\bf a}$) or $abba$  (if the links are repeated, which contributes with a term without $W_{\bf a}$), see Fig.~\ref{fig:FeynmanDiagrams2}.

On the other hand, the remaining three fourth-order terms appearing in Eq.~(\ref{Heff_resonant}), with their Hermitian conjugates, are disconnected. To see this, we note that at most one of the four operators, the $T_{\pm 2}$, can act on a $y$-link, while the other three, of the form $T_n^\pm$, only contain $x$-link terms. Therefore, the only possibility is that two pairs of hard-core bosons are created on distinct $x$ links and annihilated on the same two links, a processes that can be represented with two small disconnected loops $aa$ and $bb$, see Fig.~\ref{fig:FeynmanDiagrams2}. These processes can contribute with terms which are of order $N^2$ instead of $N$. Clearly, this does not occur for the read-out schemes discussed in Sec.~\ref{sec:readout}, where each cavity mode is coupled locally to a single plaquette or star, but is relevant for Sec.~\ref{sec:resonant_enhancement}. By assuming the homogeneous coupling scheme (\ref{couplings_enhanced_gap}), the disconnected terms of Eq.~(\ref{Heff_resonant}) evaluate to
\begin{eqnarray}\label{eq:disconnectedterms}
&-&\frac12 \frac{\delta_x^3(N^2-3N)}{(4J_{z}-\omega_{1})^3}(J_{x}a^{\dagger2}a+\delta_{x}a^{\dagger2}a^2+\delta_{x} a^{\dagger2}a a^{\dagger}+\m{H.c.})\nonumber\\
&+&\frac{5}{8}\frac{\delta_x^3N^2}{(4J_{z}-\omega_{1})^3}(J_{x}a^{\dagger}aa^{\dagger}+\delta_{x}a^{\dagger}aa^{\dagger}a+\delta_{x}a^{\dagger}aa^{\dagger2}+\m{H.c.})\nonumber\\
&-&\frac{1}{8}\frac{\delta_x^3N^2}{(4J{z}-\omega_{1})^3}(J_{x}a^{\dagger2}a+\delta_{x}a^{\dagger}aa^{\dagger}a+\delta_{x}a^{\dagger3}a+\m{H.c.}). \nonumber\\
\end{eqnarray}
It is immediate to check that, if the photon operators are replaced by complex numbers, i.e., $a^{(\dag)}\to \alpha^{(*)}$, Eq.~(\ref{eq:disconnectedterms}) gives an extensive contribution since all the $N^2$ terms cancel, in agreement with the linked-cluster theorem. If we use the bosonic commutation relation $\left[a,a^{\dagger}\right]=1$, then we can simplify Eq.~(\ref{eq:disconnectedterms}) to the following form
\begin{eqnarray}\label{eq:N2terms}
&&-\frac{3N}{2}\delta_x^3[\delta_{x}( a^{\dagger2}a^2+a^{\dagger2}a a^{\dagger}+\m{H.c.})+  J_{x}(a^{\dagger2}a+a^{\dagger}a^2)]\nonumber\\
&&+\frac{N^2}{8}\delta_x^3 [8 \delta_{x}a^{\dagger}a+5 J_{x}(a+a^{\dagger})+6 \delta_{x}(a^2+a^{\dagger2})],
\end{eqnarray}
where the $N^2$ contribution (second line) does not vanish.

This result does not represent a fundamental problem since, for a finite system, terms proportional to $N^2$ or higher powers of $N$ can certainly exist. However, the final perturbative expressions are not extensive quantities in the thermodynamic limit $N \to \infty$. This behavior can be attributed to the failure of the perturbative treatment since, besides the conditions $J_{x,y},\,\delta_{x}\sqrt{\langle a^\dag a \rangle}\ll (4J_{z}-\omega)$, the requirement to have a power series in terms of small expansion parameters also leads to an upper bound on the size of $N$, namely
\begin{eqnarray}\label{N_disconnected_bound}
\delta_{x}\sqrt{N}\ll(4J_{z}-\omega_{x}).
\end{eqnarray}
This condition is not very restrictive, since the coupling $\delta_x$ is expected to be small. Furthermore, an interesting limit is obtained for a vanishing coupling $\delta_x$ and a large number of photons $\langle a^\dag a \rangle$. In this case, for system size $N \ll \langle a^\dag a \rangle $, the non-extensive terms can be generally neglected in comparison to the extensive ones. For example, in Eq.~(\ref{eq:N2terms}) this condition would give $\delta_x^3 N \langle a^\dag a\rangle^2 \gg \delta_x^3 N^2 \langle a^\dag a\rangle $, by comparing the first and second line.

\section{SW TRANSFORMATION IN THE PRESENCE OF FOUR DIFFERENT LINKS AND TWO RESONANT CAVITY MODES}\label{app:SW2cavities}

In this section we present how the SW transformation is applied to the scheme of Sec.~\ref{sec:long-range} for which the four links belonging to each $W_{\bf a}$ operator have all distinct couplings, see Eqs.~(\ref{half_coupling_x}), (\ref{half_coupling_y}), and Fig.~\ref{fig:differentlinks}. By using Eqs.~(\ref{SW_2}) and (\ref{SW_4}) one can obtain the zero hard-core boson effective Hamiltonian up to fourth order. Since we are interested in a regime where $(4J_{z}-\omega_{x,y})\ll\omega_{x,y}$, we keep the resonant terms only. Furthermore, it is sufficient to keep at fourth order only the terms that couple to the star and plaquette operators. This leads to:
\begin{eqnarray}\label{eq:HamiltonianT}
H_{\m{eff}}=&&\sum_{\alpha=x,y}\left[\omega_{\alpha}a_{\alpha}^{\dagger}a_{\alpha}
-\frac{T_{-2}^{\alpha}(T_{+2}^{0}+T_{+2}^{\alpha}+T_{+2}^{-\alpha})+\m{H.c.}}{2(4J_{z}-\omega_{\alpha})} \right.\nonumber\\
&& \left. -\frac{T_{-2}^{+\alpha}T_{0}^{0}T_{0}^{0}(T_{+2}^{0}+T_{+2}^{-\bar\alpha})+\m{H.c.}}{2(4J_{z}-\omega_{\alpha})^3}\right].
\end{eqnarray}
where $\bar\alpha=y$ if $\alpha=x$ and vice versa. In the above expression, terms like $T_{-2}^{\alpha}T_{+2}^{\beta}T_{-2}^{\gamma}T_{+2}^{\zeta}$ or $T_{-2}^{\alpha}T_{-2}^{\beta}T_{+2}^{\gamma}T_{+2}^{\zeta}$ do not appear, because they do not contribute to the resonant coupling to the $W_{\bf a}$ operators. It is worth mentioning that the conditions in order to treat the Hamiltonian within perturbation theory are the same as in Appendix \ref{app:SWONEMODE}, namely,
\begin{equation}\label{conditions_simple}
J_{\alpha},~\delta_{\alpha} \sqrt{\langle a_\alpha^\dag a_\alpha \rangle},~\delta_\alpha\sqrt{N}  \ll \min ( 4J_{z}-\omega_{x,y}).
\end{equation}
By writing the Hamiltonian (\ref{eq:HamiltonianT}) explicitly in terms of effective spin operators and photon operators, we obtain the following zero hard-core boson effective Hamiltonian
\begin{eqnarray}\label{H_photons_2modes}
H_{\m{eff}}&&=\sum_{\alpha=x,y}\Bigg[\omega^\prime_{\alpha}a_{\alpha}^{\dagger}a_{\alpha}
-\frac{J_{\alpha}\delta_{\alpha}}{4J_{z}-\omega_{\alpha}}\left(a_{\alpha}+a_{\alpha}^{\dagger}\right)\frac{N}{4}\nonumber\\
&&-\frac{\delta_{\alpha}^2}{4J_{z}-\omega_{\alpha}}\left(a_{\alpha}^2+a_{\alpha}^{\dagger2}\right)\frac{N}{4}\nonumber\\
&&-\frac{J_{\bar\alpha}^2 J_{\alpha}\delta_\alpha}{(4J_{z}-\omega_{\alpha})^3}\left(a_{\alpha}+a_{\alpha}^{\dagger}\right)\sum_{\bf a}W_{\bf a}\nonumber\\
&&-\frac{J_xJ_y\delta_{x}\delta_{y}}{2(4J_{z}-\omega_{\alpha})^3}\left(a_{x}^{\dagger}a_{y}+a_{x}a_{y}^{\dagger}\right)\sum_{\bf a}W_{\bf a}\Bigg].
\end{eqnarray}
In the strictly resonant case [see Eq.~(\ref{condition_very_resonant})] we can drop all the nondiagonal terms except the last line, which leads to Eq.~(\ref{eq:firstSWresonant}). If we want to calculate the leading nonresonant correction to the chemical potential, we have to keep also the $(a_\alpha+a_\alpha^\dag)$ terms in the first and third line, which leads to Eq.~(\ref{eq:firstSWnotsoresonant}). This is because, as explained in Sec.~\ref{off_resonant_gap}, these two nondiagonal terms combine to a higher-order $\sum_{\bf a} W_{\bf a}$ contribution, diagonal in the photon modes, after a second SW transformation. On the other hand, $(a_{\alpha}^2+a_{\alpha}^{\dagger2})$ does not appear together with $\sum_{\bf a} W_{\bf a}$ and therefore can be dropped, for the purpose of obtaining corrections to the star and plaquette couplings. Finally we note that the third line, after the second SW transformation in the photon operators, also leads to a interaction term $(\sum_{\bf a} W_{\bf a})^2$, but this is generally negligible compared to the one from the last line of Eq.~(\ref{H_photons_2modes}) and is not considered in Sec.~\ref{off_resonant_gap}.

As a last note, a conservative requirement to perform the SW on Eq.~(\ref{H_photons_2modes}) is that that the off-diagonal operators are much smaller than the corresponding gaps. From the first line of Eq.~(\ref{H_photons_2modes}) we obtain
\begin{equation}\label{eq:consition1N}
\frac{J_{\alpha}\delta_{\alpha}\sqrt{\langle a_\alpha a_\alpha^\dag \rangle}}{4J_{z}-\omega_{\alpha}}\frac{N}{4}\ll\omega_\alpha',
\end{equation}
and from the fourth line
\begin{equation}
\frac{J_x J_y\delta_{x}\delta_{y}\sqrt{\langle a_x a_x^\dag \rangle\langle a_y a_y^\dag \rangle}}{2(4J_{z}-\omega_{\alpha})^3}N\ll|\omega_x'-\omega_y'|.
\end{equation}
Finally, the condition from the third line of Eq.~(\ref{H_photons_2modes}) is less restrictive than Eq.~(\ref{eq:consition1N}), because of an additional small factor $J_{\bar\alpha}^2/(4J_z-\omega_\alpha)^2$. These conditions, together with Eq.~(\ref{conditions_simple}), pose a restriction onto the size $N$ at which the perturbative treatment is justified.

\section{SW TRANSFORMATION IN THE PRESENCE OF SMALL FREQUENCY MODES}\label{app:offresonant}

The main purpose of this section is to prove the prescription (\ref{prescription_non_res}) of Sec.~\ref{sec_not_resonant} to obtain the effective Hamiltonian in the presence of off-resonant cavity modes.

First we note that, if no cavity is present, only the $T^0_n$ operators are nonvanishing and $T_n=T_n^0$. Therefore, we can drop in Eqs.~(\ref{SW_1}-\ref{SW_4}) the summations in the photon indexes $i,j,k,r$ and set all the $T^i_n$ operators equal to $T_n$. On the other hand, by making use of Eq.~(\ref{small_frequency}) for the case with small frequency modes, we can approximate $\delta\epsilon_n^i \simeq 2nJ_z$, i.e., neglect the photon frequency shifts in the denominators of Eqs.~(\ref{SW_1}-\ref{SW_4}). This allows to perform the summations in the photon indexes and express Eqs.~(\ref{SW_1}-\ref{SW_4}) in terms of the $T_n$ operators, as for the case without photon modes discussed above. The difference is that the $T_n$ are now photon operators, simply obtained from the ones without cavities by substituting $J_{{\bf a},k}\to K_{{\bf a},k}$ (where $k=x,y$). Furthermore, it is also clear from Eq.~(\ref{eq:couplings}) that the $K_{{\bf a},k}$ are all commuting operators, and therefore they can be treated in the same way of the $J_{{\bf a},k}$ in the derivation of the effective Hamiltonian. This shows that it is sufficient to apply Eq.~(\ref{prescription_non_res}), if the effective Hamiltonian without cavity modes is known.

By applying this prescription to the second coupling scheme of Sec.~\ref{sec_not_resonant} [see Eq.~(\ref{interacting_scheme_offres})], we obtain the following effective Hamiltonian:
\begin{eqnarray}\label{off_res_photon_H}
&&H_{\m{eff}}=\sum_{\alpha=x,y}\Bigg[\omega^{\prime\prime}_{\alpha}a_{\alpha}^{\dagger}a_{\alpha}
-\frac{J_{\alpha}\delta_{\alpha}}{4J_{z}}(a_{\alpha}+a_{\alpha}^{\dagger})N\nonumber\\
&&-\frac{\delta_{\alpha}^2}{4J_{z}}(a_{\alpha}^2+a_{\alpha}^{\dagger2})\frac{N}{2}
-\frac{J_{\bar\alpha}^2J_\alpha \delta_{\alpha}}{16J_{z}^3}(a_{\alpha}+a_{\alpha}^{\dagger})\sum_{\bf a}W_{\bf a}\Bigg]\nonumber\\
&&-\frac{J_{x}^2J_{y}^2}{16J_{z}^3}\sum_{\bf a}W_{\bf a}-\frac{J_{x}J_{y}\delta_{x}\delta_{y}}{16J_{z}^3}
\left(a_{x}a_{y}^{\dagger} +a_{y}^{\dagger}a_{x}\right)\sum_{\bf a}W_{\bf a}\nonumber\\
&&-\frac{J_{x}J_{y}\delta_{x}\delta_{y}}{16J_{z}^3}
\left(a_{x}a_{y}+a_{x}^{\dagger}a_{y}^{\dagger}\right)\sum_{\bf a}W_{\bf a},
\end{eqnarray}
where $\omega^{\prime\prime}_\alpha$ are defined in Eq.~(\ref{omega_double_prime}). In Eq.~(\ref{off_res_photon_H}), we only kept terms from the first and last lines of Eq.~(\ref{eq:SWTtwocouplings}) while dropping all the fourth order terms which do not couple to the $W_{\bf a}$ operators and some inessential constants.

We note now that the $(a_{\alpha}^2+a_{\alpha}^{\dagger2})$ term, in the second line of Eq.~(\ref{off_res_photon_H}), does not have a corresponding $\sum_{\bf a} W_{\bf a}$ term, and that the off-resonant coupling to the $W_{\bf a}$ operators in the last line does not have a corresponding $\left(a_{x}a_{y}+a_{x}^{\dagger}a_{y}^{\dagger}\right)$ term at lower order. Therefore, after a second SW transformation of Eq.~(\ref{off_res_photon_H}), these two terms give negligible corrections to the effective Hamiltonian for the star and plaquette operators and we have dropped them in Eq.~(\ref{Heff_nonres_2cavities}).



\begin{thebibliography}{100}
\bibitem{self_energy_comment}
We also include in the interaction term of Eq.~(\ref{interacting_anyons}) a self-energy correction $A/2$.
\end{thebibliography}
\end{document}